\documentclass[]{raa}            
\usepackage{graphicx,times}
\usepackage{natbib}
\usepackage{rotating}
\usepackage{threeparttable}

\providecommand{\bm}[1]{\mbox{\boldmath$#1$\unboldmath}}

\begin{document}

\title{M dwarf Stars --The By-Product of X-Ray Selected AGN Candidates
}

\volnopage{ {\bf 2012} Vol.\ {\bf 0} No. {\bf XX}, 000--000}
\setcounter{page}{1}

\author{Y. Bai\inst{1}, Y.-C. Sun\inst{1}, X.-T. He\inst{1},
Y. Chen\inst{1}, J.-H. Wu\inst{1}, Q.-K. Li\inst{1},
R.-F. Green\inst{2} \& W. Voges\inst{3}}

\institute{Department of Astronomy, Beijing Normal University,
Beijing 100875, China; {\it baiyu@bnu.edu.cn}\\
\and
Kitt Peak National Observatory, NOAO, Tucson, AZ85726-6732, USA\\
\and
Max-Planck-Institute f\"ur Extraterrestrische Physik, D-85740 Garching, Germany\\
}
\vs \no
   {\small Received [year] [month] [day]; accepted [year] [month] [day] }

\abstract{
X-ray loud M dwarfs are a major source of by-product (contamination)
in the X-ray band of the multiwavelength quasar survey (MWQS). As a
by-product, the low dispersion spectra of 22 M dwarfs are
obtained in which the spectra of 16 sources are taken for the first time.
The spectral types and distance of the sample are given based
on spectral indices CaH2, CaH3, and TiO5. The parameter
$\zeta_{TiO/CaH}$ is calculated to make the
metallicity class separation among dwarfs, subdwarfs and
extreme subdwarfs. We also discuss the distributions in the
diagrams of Log($L_{x}/L_{bol}$) versus spectral type and
infrared colors.
}

   \authorrunning{}            
   \titlerunning{}  
   \maketitle

\section{Introduction}

Observations from X-ray satellites have
showed that strong X-ray emission is a quasi-defining
characteristics of Active Galactic Nuclei (AGNs),
and X-ray selection from a deep X-ray survey might
be the best census method to obtain a sample of AGNs
with a high completeness level (\cite{He01}). The
multiwavelength quasar survey(MWQS) aimed at obtaining
a sample of $\sim$350 new AGNs brighter than B = 19.0,
based on a simple yet effective X-ray, radio and
optical inspection criteria (\cite{He01}; \cite{Chen02,Chen07};
\cite{Bai07}). In X-ray band survey, AGN candidates are
selected from the {\it ROSAT} All-Sky Survey (RASS)
sources by applying the criteria to exclude most
stars and galaxy clusters (\cite{Chen07}). However,
M dwarfs are hard to be distinguished from AGN
candidates if criteria in soft X-ray band alone
are applied, thus constituting a potential contaminant
to the AGN candidates in optical observations.
By analyzing low dispersion spectral characteristics
of AGNs and M dwarfs, these two kinds of sources
can be differentiated.  Thus the selected X-ray loud
M dwarfs can be investigated in details.

M dwarf stars are the most numerous stars in our
galactic neighborhood with masses between 0.1 and
0.5 $M_{\odot}$, and they can be divided into two
classes (dM and dMe) based on an optical
criterion: dMe stars have $H_{\alpha}$ in
emission, while dM stars have $H_{\alpha}$ in
absorption. Due to the large quantity, longevity
(with lifetime greater than the Hubble time), and
spectral energy distribution (SED) sensitivity to
metallicity (\cite{Lepine07}), M dwarfs are suitable
as tracers of the chemical and dynamical evolution
of the Galaxy (\cite{Laughlin97}). The optical and
near infrared spectra of M dwarfs are dominated by
molecular absorption bands of metal oxides and hydrides
(\cite{Bessel91}) and the ratio between the
strength of them has been known as a metallicity
diagnostic (\cite{Bessel82}).Four spectroscopic
indices (TiO5, CaH1, CaH2, CaH3) was defined
by \cite{Reid95} to make a separation of three
metallicity subclasses, M
dwarfs (dM) subdwarfs (sdM) and extreme subdwarfs
(esdM) (\cite{Gizis97}). Furthermore, the higher
radial velocity and deeper transit/occultation as
the consequence of the low stellar mass and small
radius of M dwarf stars will facilitate the detection
of small planets.  M dwarfs might be suitable for
hosting planets, as indicated by the common observed
accretion disks around young dwarf stars (\cite{Liu04};
\cite{Lada06}).

In this paper, the initial result of a sample of
22 M dwarfs limited in X-ray flux is given in Section 2.
In Section 3, the classification system for M dwarfs
and subdwarfs is re-examined based on a combinations
of four relationships defined in [CaH1, TiO5], [CaH2,
TiO5], and[CaH3, TiO5]. The metallicity index
$\zeta_{TiO/CaH}$ is calculated to measurement
the metal content in M dwarfs. A summary is given in
Section 4.

\section{Observation and result}

The candidates are selected with criteria in X-ray band
and observed by using 2.1m and 2.16m telescopes at
KPNO (Kitt Peak National Observatory) and NAOC
(National Astronomical Observatories, Chinese
Academy of Sciences) (\cite{He01, Chen02,
Chen07, Bai07, Sun11}). The criteria adopted for
the selection of candidates are consistent
between two telescopes(\cite{Bai07}).
The spectral coverages are respectively about
4000-10000{\AA} with resolution of 7{\AA} and
4000-8000{\AA} with resolution of 13{\AA}. All
the spectra are reduced with IRAF.
The initial result is a subset of 22 M dwarfs,
in which the spectra of 16 sources are taken for the first time.
A cross identification has been made with the 2MASS
(Two Micron All Sky Survey) All-Sky Catalog of Point
Sources to get data in the band of near infrared.
All of 2MASS sources lie within 7" to the optical
position of our targets. Table 1 displays the ROSAT
and 2MASS source designation and optical position.
Among 22 M dwarfs, four spectra have $H_{\alpha}$ in
absorption and are classified as dM stars
(1RXS J015515.4-173916, 1RXS J041615.7+012640,
1RXS J043426.2-030041, 1RXS J075107.8+061714).
Others are classified as dMe stars for the
$H_{\alpha}$ in emission. Fig.~\ref{spectra}
shows the spectra of the M dwarfs.

\begin{table}[h]

\centering

\caption{Position of M dwarfs.}

\tabcolsep 6mm
 \begin{tabular}{lllr}
  \hline\noalign{\smallskip}
ROSAT name            & 2MASS name       & R.A.(2000)  & DEC (2000)    \\
  \hline\noalign{\smallskip}
1RXS J015515.4-173916 & 01551663-1739080 & 01 55 16.32 & $-$17 39 06.9 \\
1RXS J040840.7-270534 & 04084031-2705473*& 04 08 40.24 & $-$27 05 46.0 \\
1RXS J041132.8+023559 & 04113190+0236058 & 04 11 31.68 &  + 02 36 05.2 \\
1RXS J041132.8+023559 & 04113149+0236012 & 04 11 31.50 &  + 02 36 01.3 \\
1RXS J041325.8-013919 & 04132663-0139211*& 04 13 26.19 & $-$01 39 20.7 \\
1RXS J041612.7-012006 & 04161320-0119554 & 04 16 13.12 & $-$01 19 54.1 \\
1RXS J041615.7+012640 & 04161502+0126570 & 04 16 14.99 &  + 01 26 57.8 \\
1RXS J042854.3+024836 & 04285400+0248060 & 04 28 53.92 &  + 02 48 08.0 \\
1RXS J043051.6-011253 & 04305207-0112471 & 04 30 51.71 & $-$01 12 48.7 \\
1RXS J043426.2-030041 & 04342887-0300390 & 04 34 28.84 & $-$03 00 39.3 \\
1RXS J053954.8-130805 & 05395494-1307598*& 05 39 54.90 & $-$13 07 59.1 \\
1RXS J055533.1-082915 & 05553254-0829243*& 05 55 32.57 & $-$08 29 25.1 \\
1RXS J060121.5-193749 & 06012092-1937421 & 06 01 20.90 & $-$19 37 42.1 \\
1RXS J075107.8+061714 & 07510760+0617135 & 07 51 07.63 &  + 06 17 13.3 \\
1RXS J111819.9+134739 & 11182030+1347392 & 11 18 20.02 &  + 13 47 41.7 \\
1RXS J112144.4+162156 & 11213897+1617528 & 11 21 39.24 &  + 16 17 52.4 \\
1RXS J125336.5+224742 & 12533626+2247354*& 12 53 36.18 &  + 22 47 35.3 \\
1RXS J130123.5+265145 & 13012236+2651422 & 13 01 22.09 &  + 26 51 46.4 \\
1RXS J170849.1-110433 & 17084975-1104308 & 17 08 50.01 & $-$11 04 25.9 \\
1RXS J174741.0-135445 & 17474148-1354368 & 17 47 41.54 & $-$13 54 35.5 \\
1RXS J185008.6+110509 & 18500888+1105098 & 18 50 08.64 &  + 11 05 10.4 \\
1RXS J210326.6+161658 & 21032686+1616569*& 21 03 26.80 &  + 16 16 57.6 \\
  \noalign{\smallskip}\hline
\end{tabular}
\tablecomments{0.86\textwidth}{$*$: These six sources
are contained in the catalog of \cite{Riaz06}. }
\end{table}

\begin{figure}
\includegraphics[width=75mm]{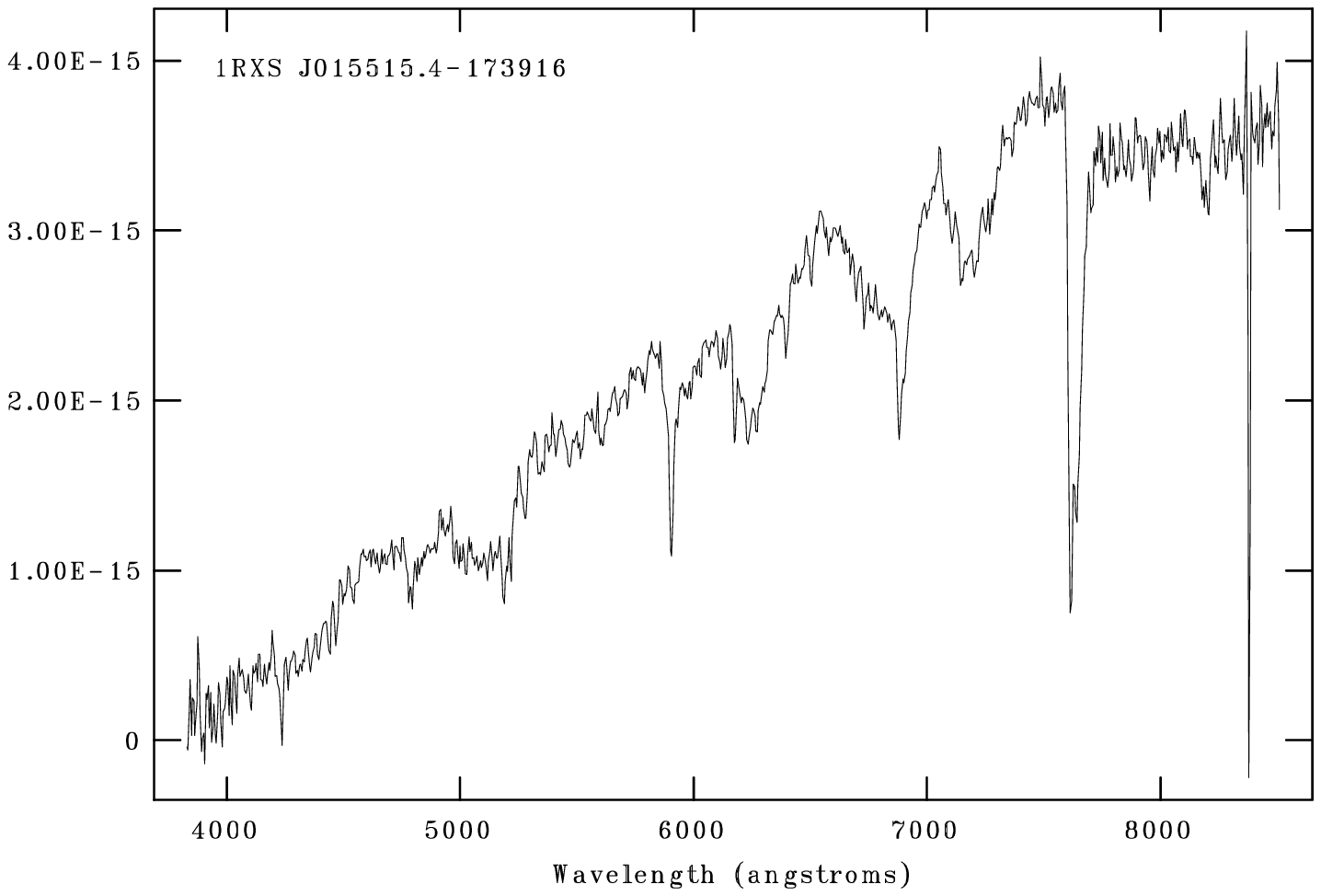}
\includegraphics[width=75mm]{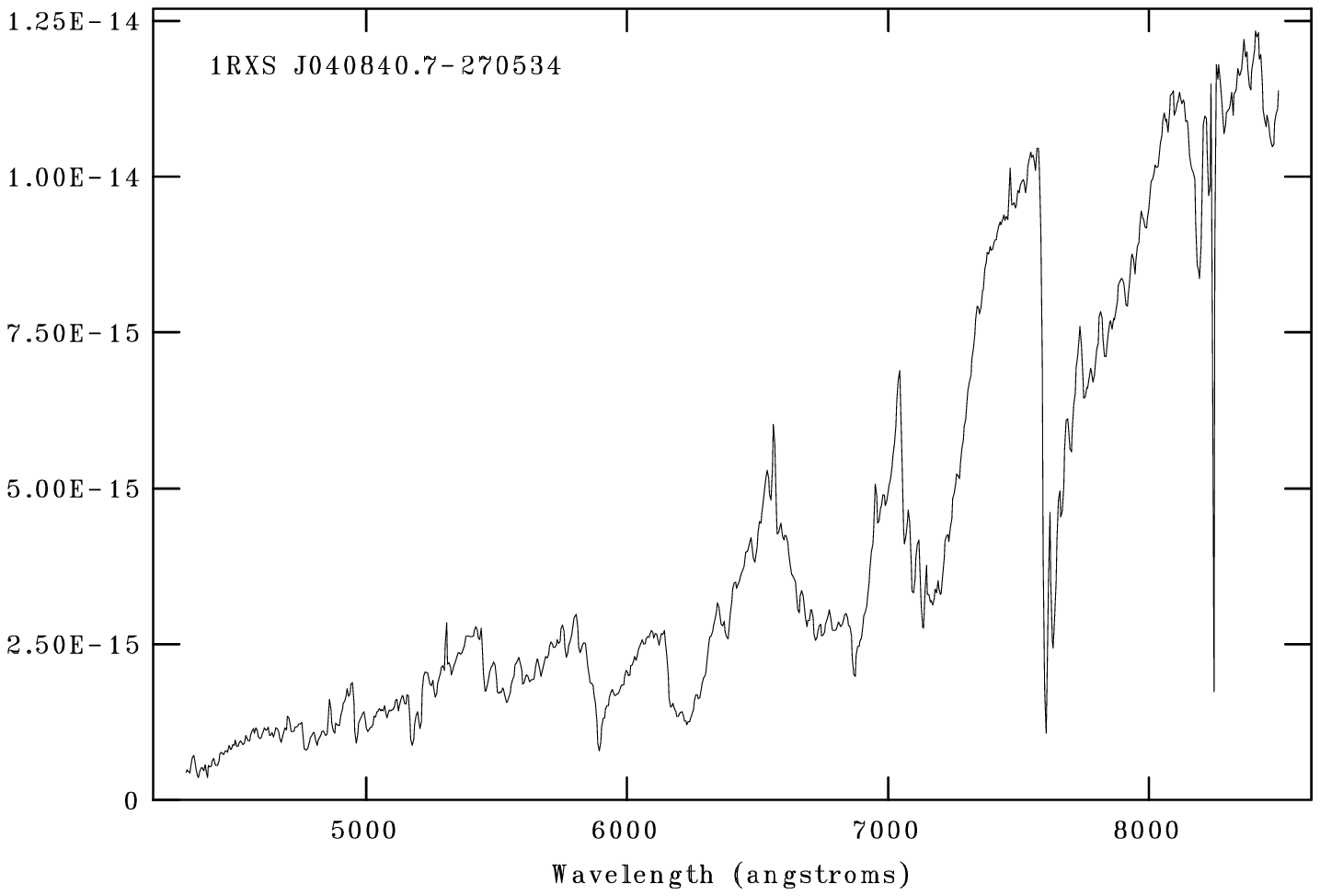}
\includegraphics[width=73mm]{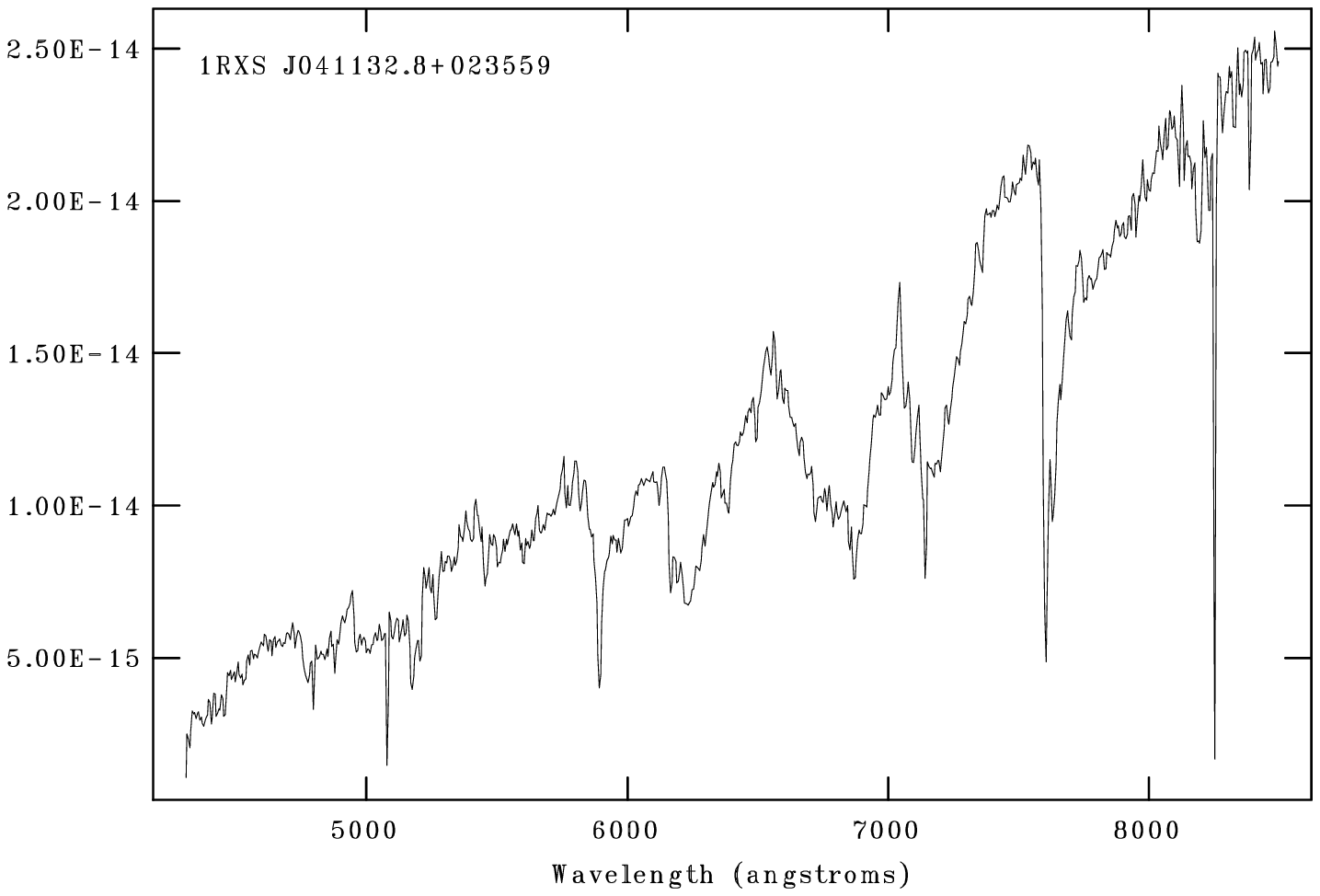}
\includegraphics[width=73mm]{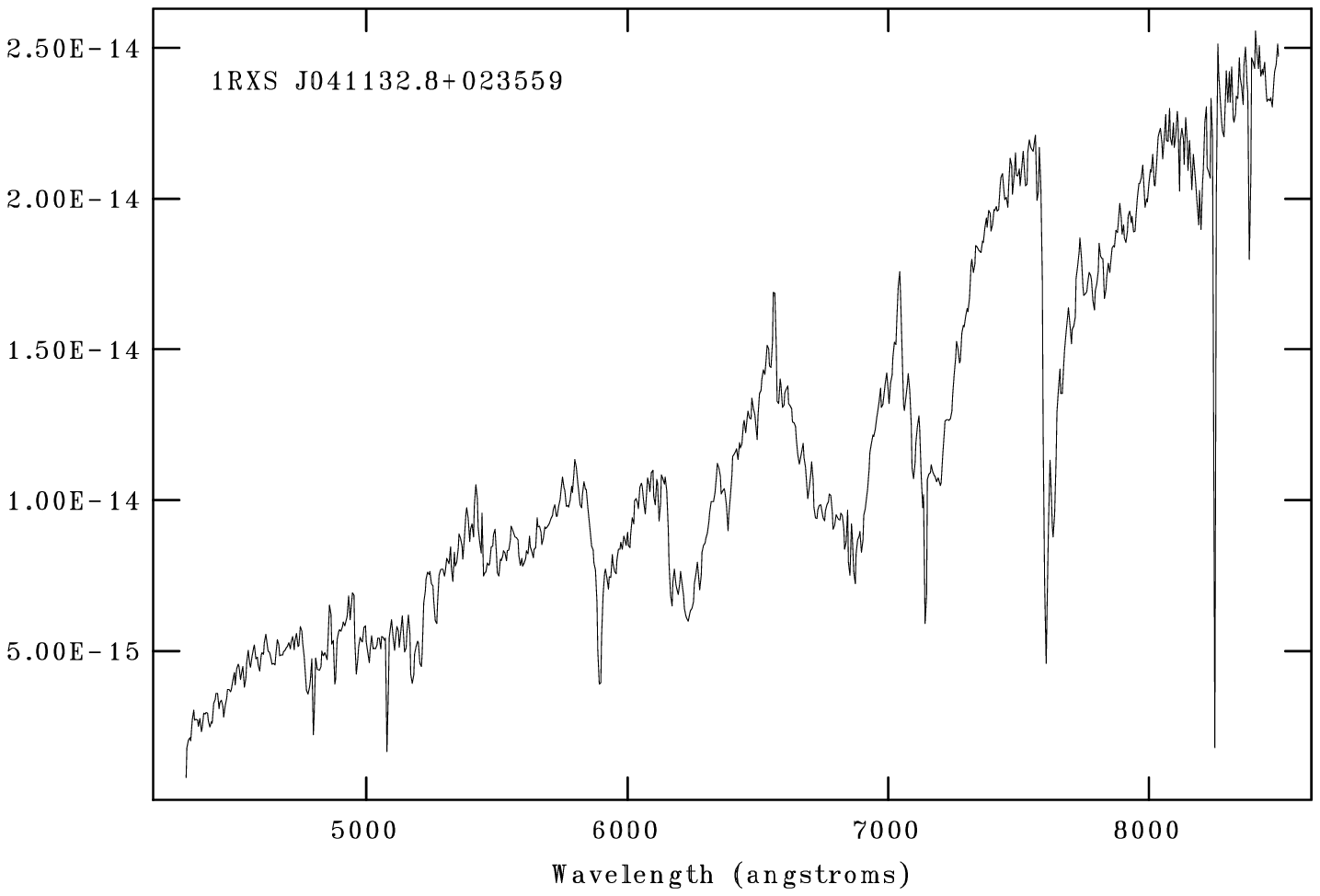}
\includegraphics[width=73mm]{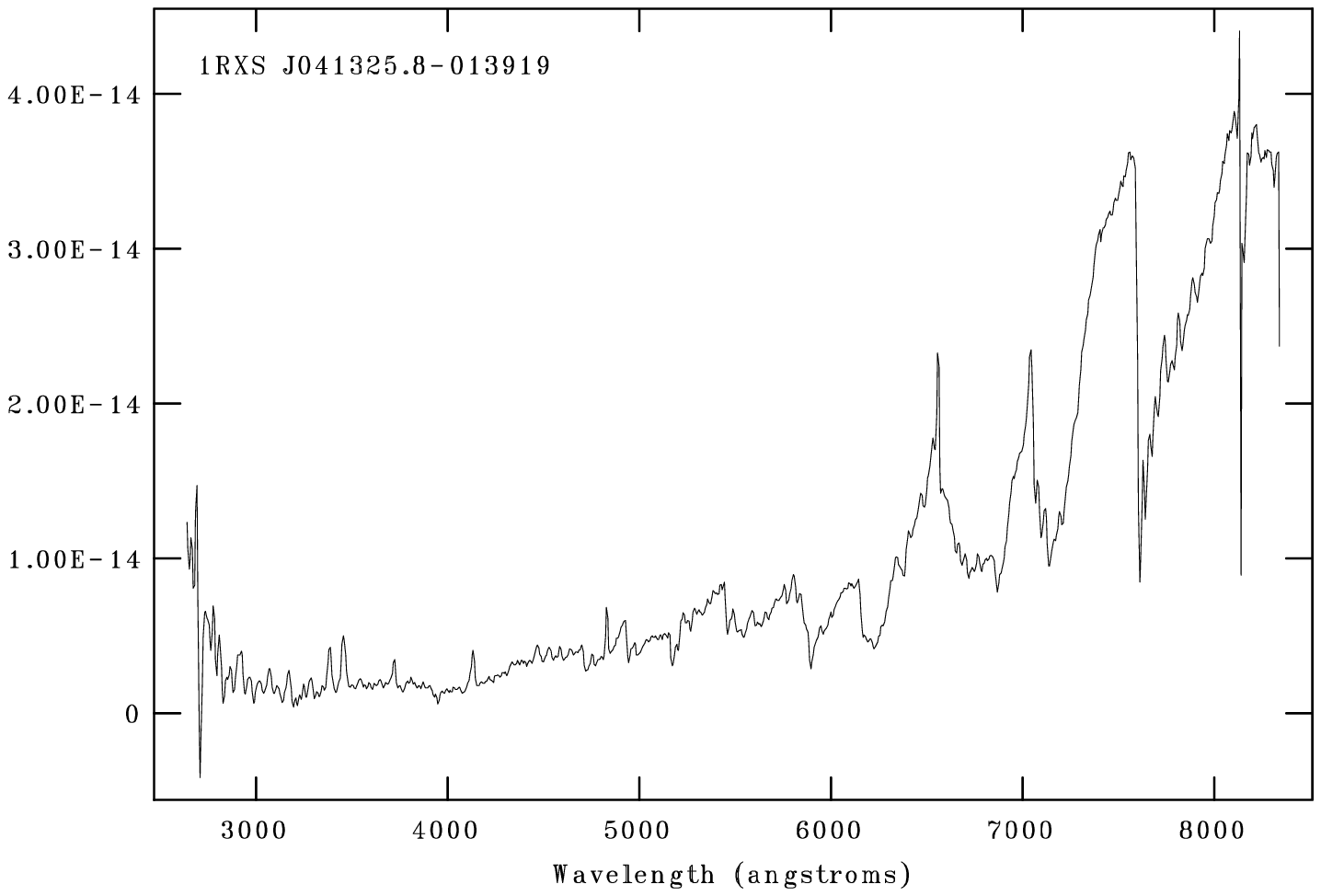}
\includegraphics[width=73mm]{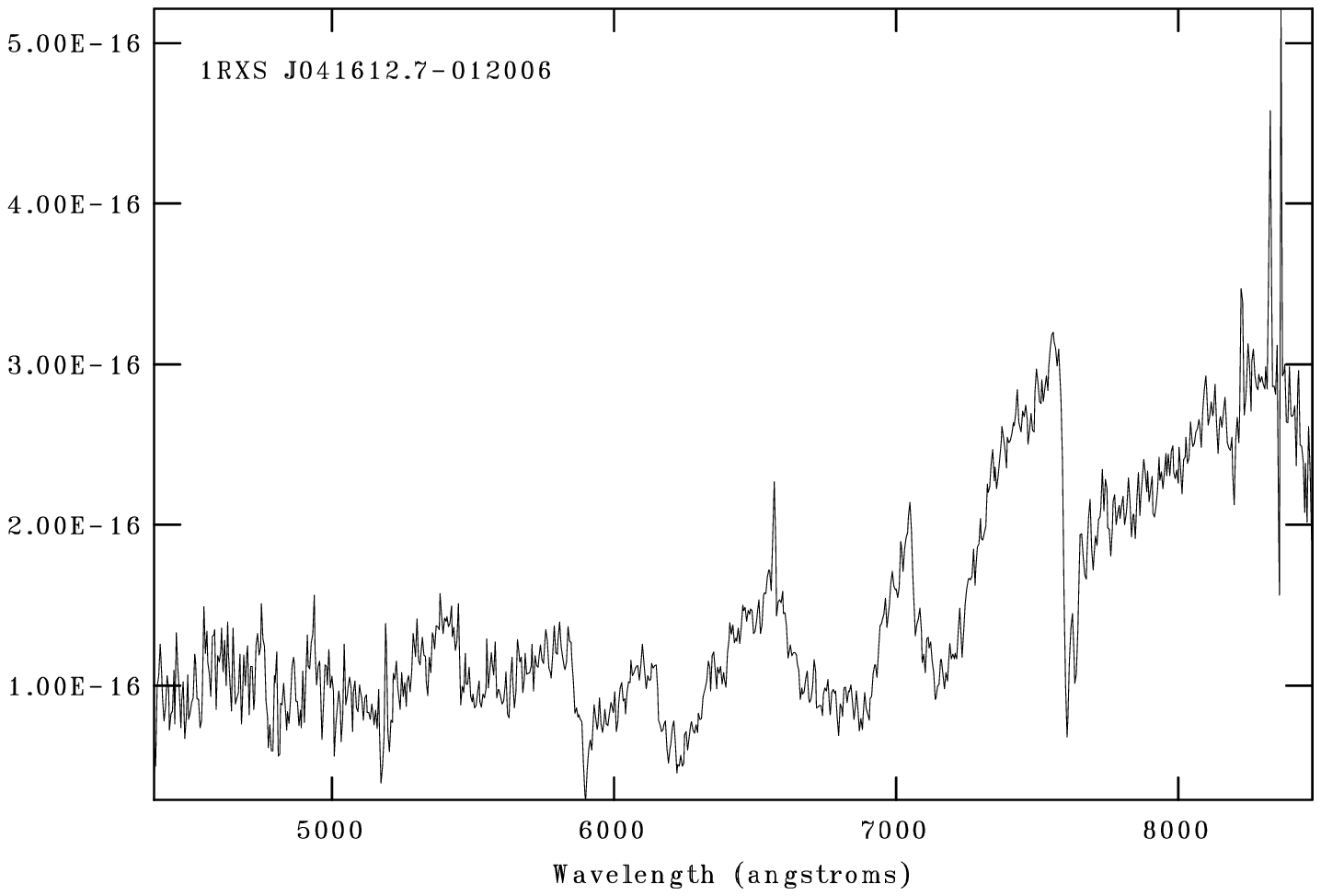}
\includegraphics[width=73mm]{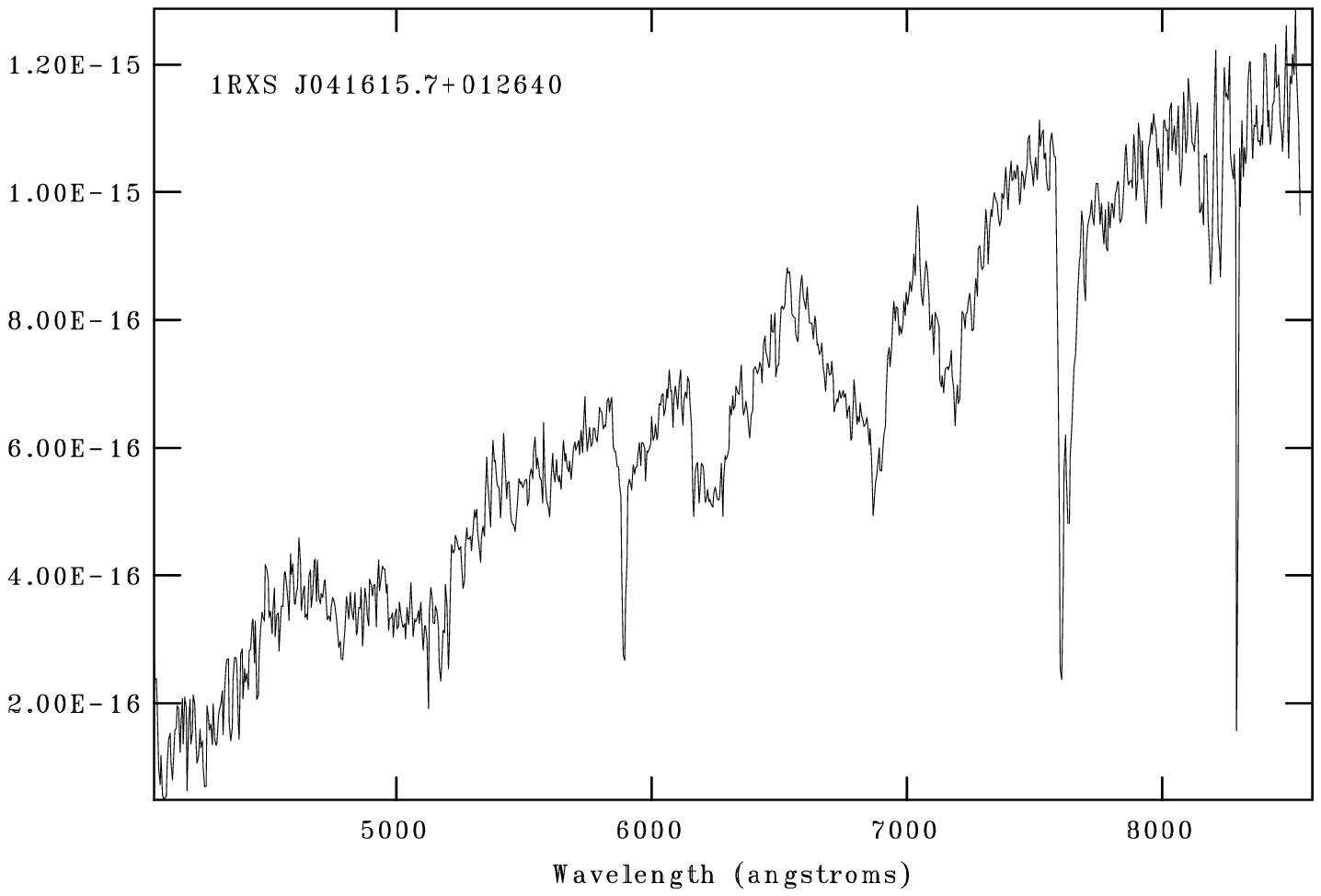}
\includegraphics[width=73mm]{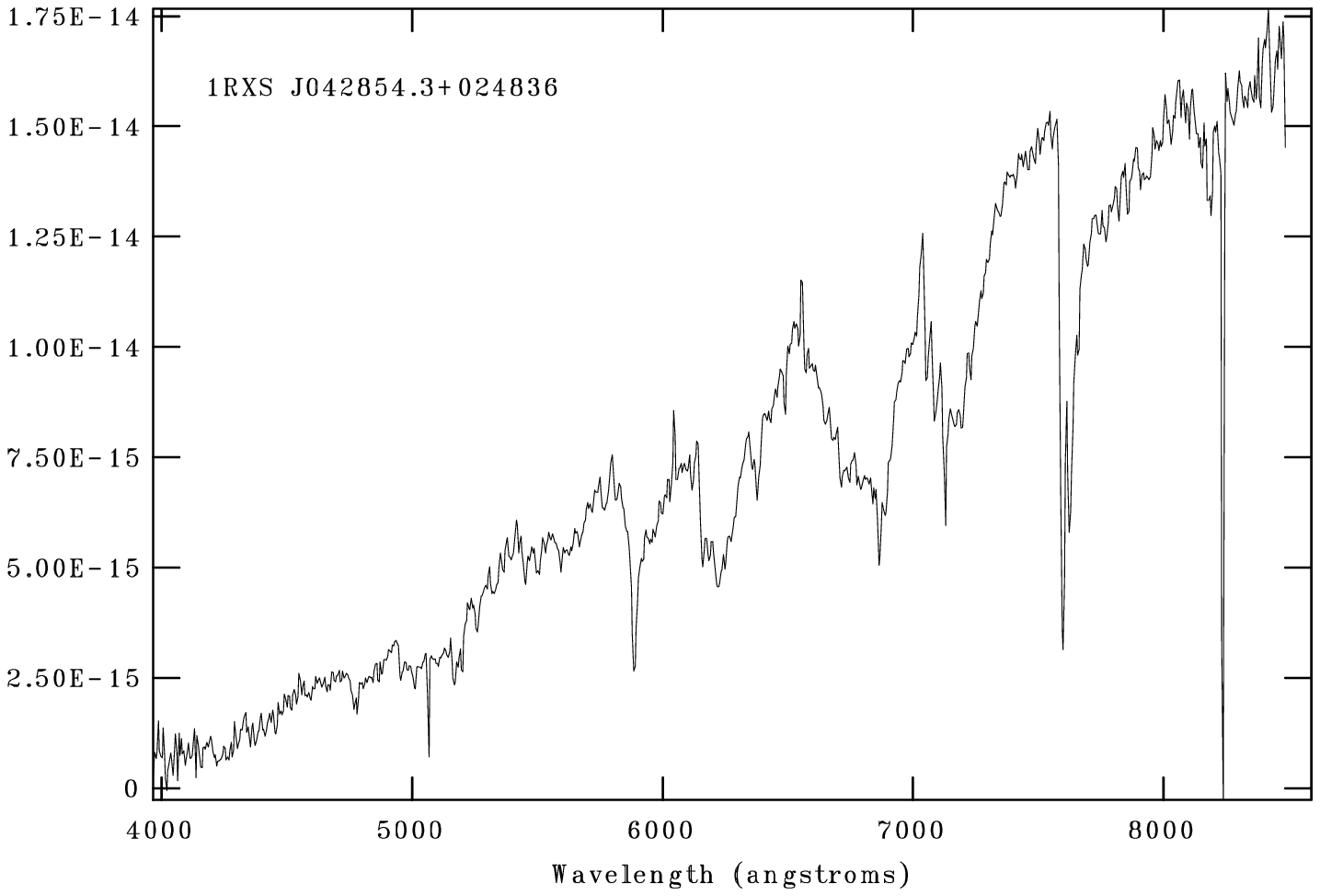}
\includegraphics[width=73mm]{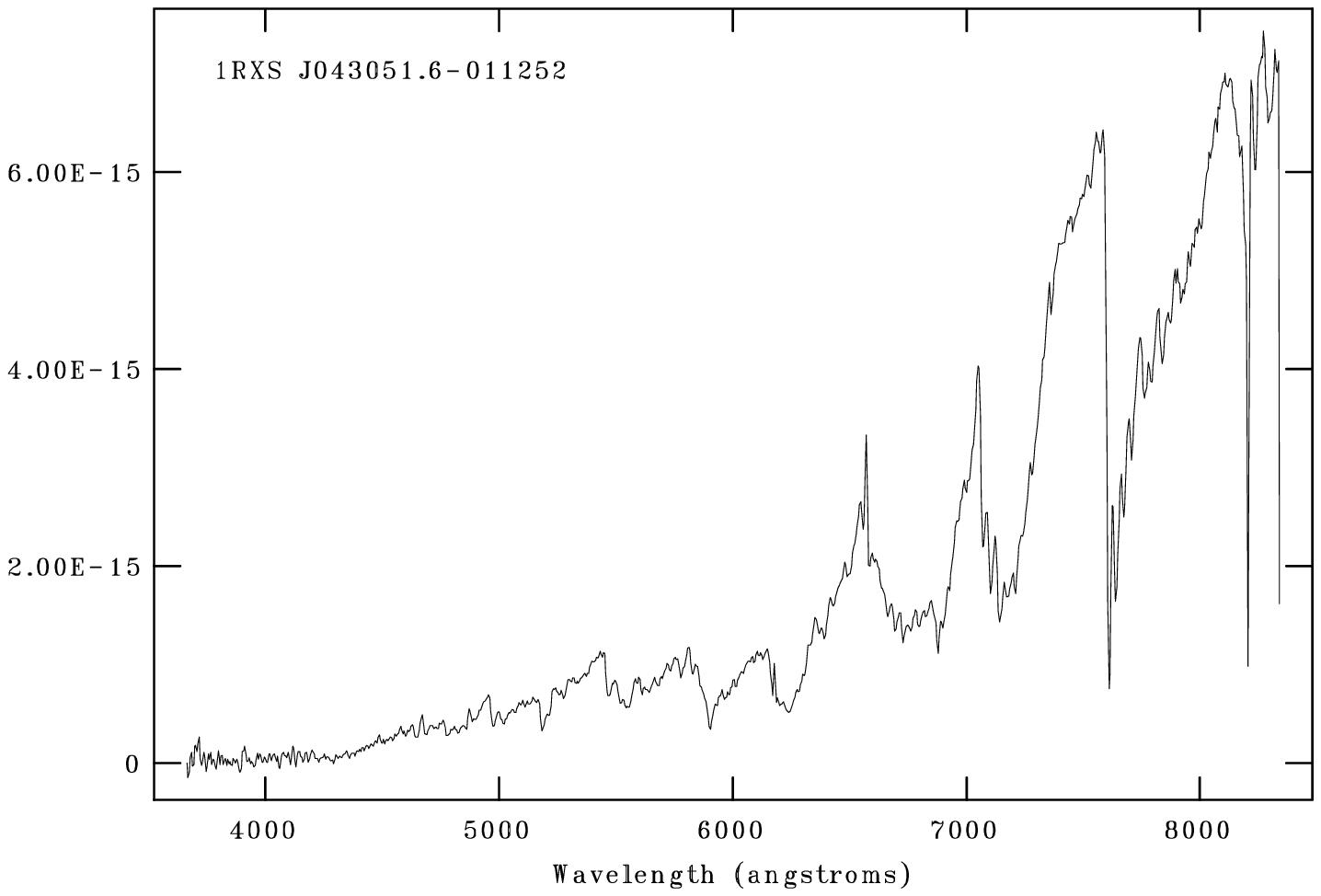}
\includegraphics[width=73mm]{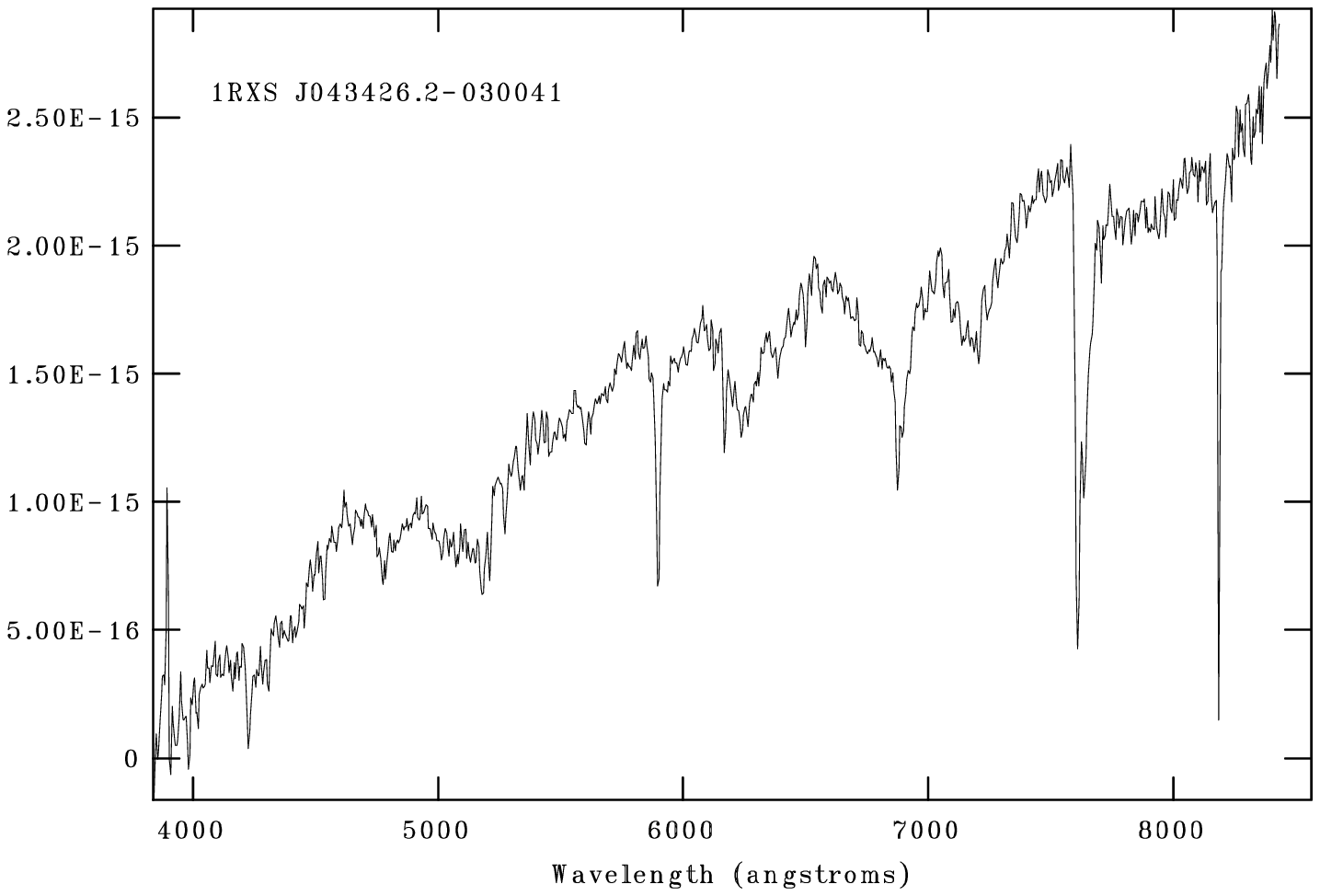}
\end{figure}
\begin{figure}
\includegraphics[width=75mm]{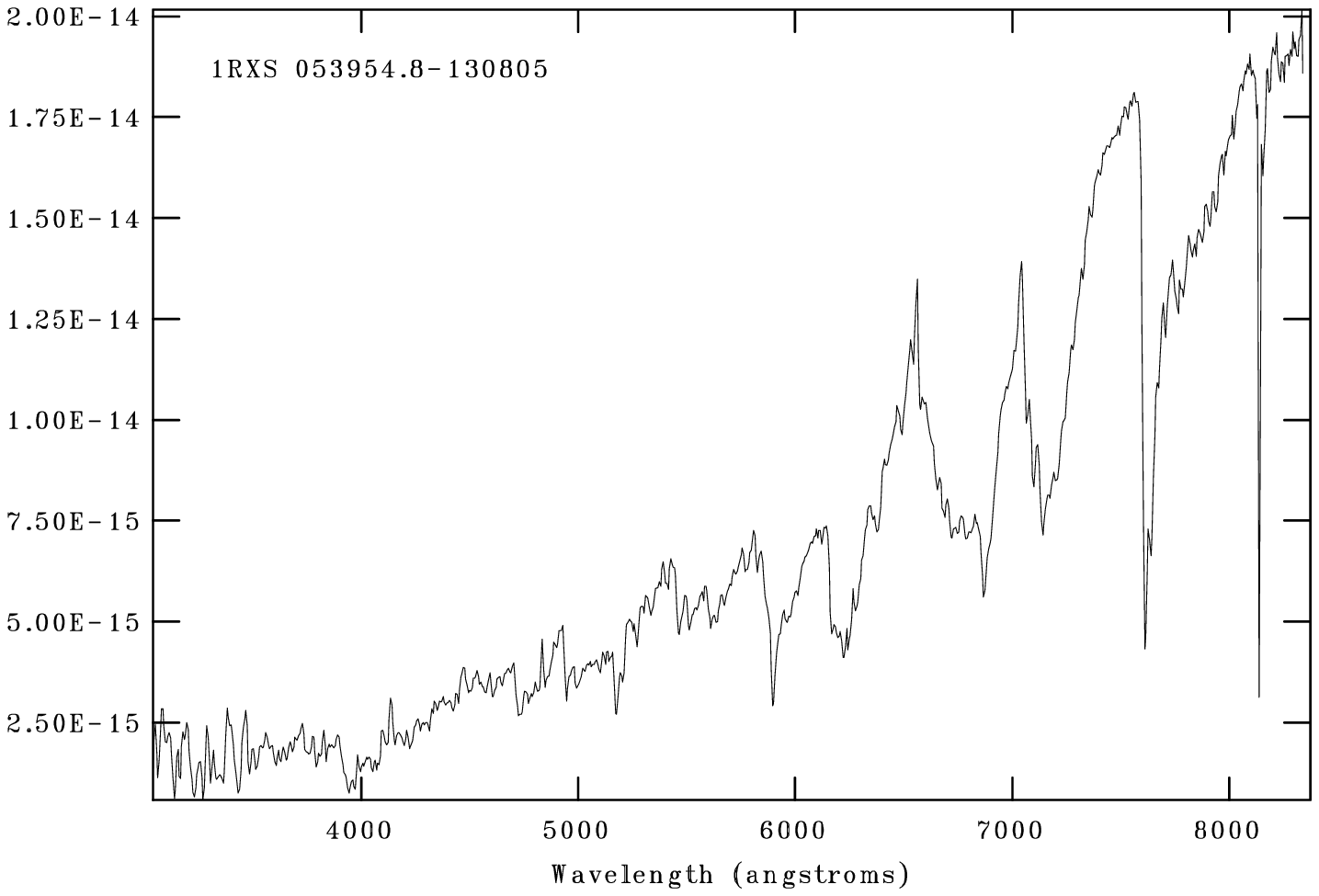}
\includegraphics[width=75mm]{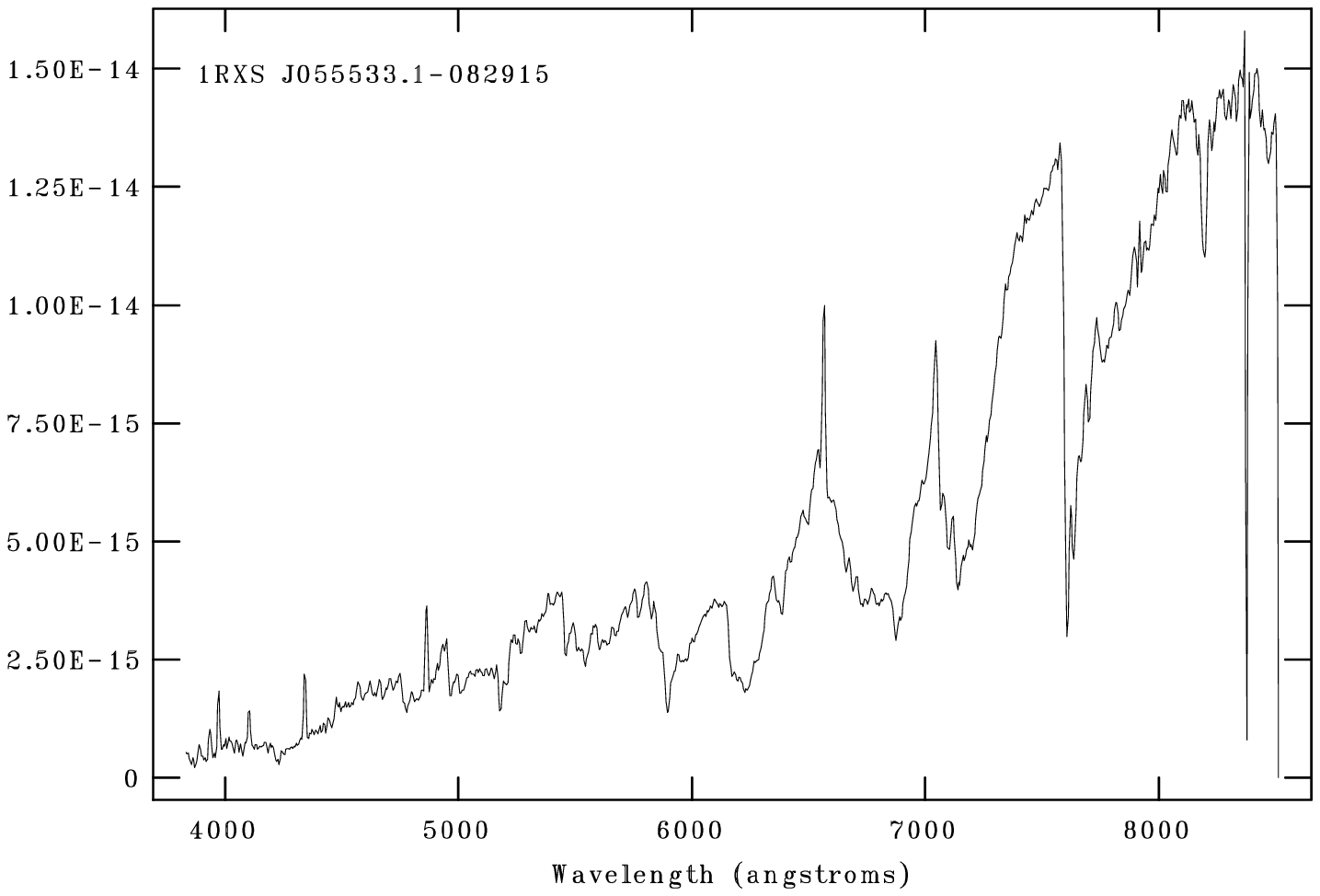}
\includegraphics[width=73mm]{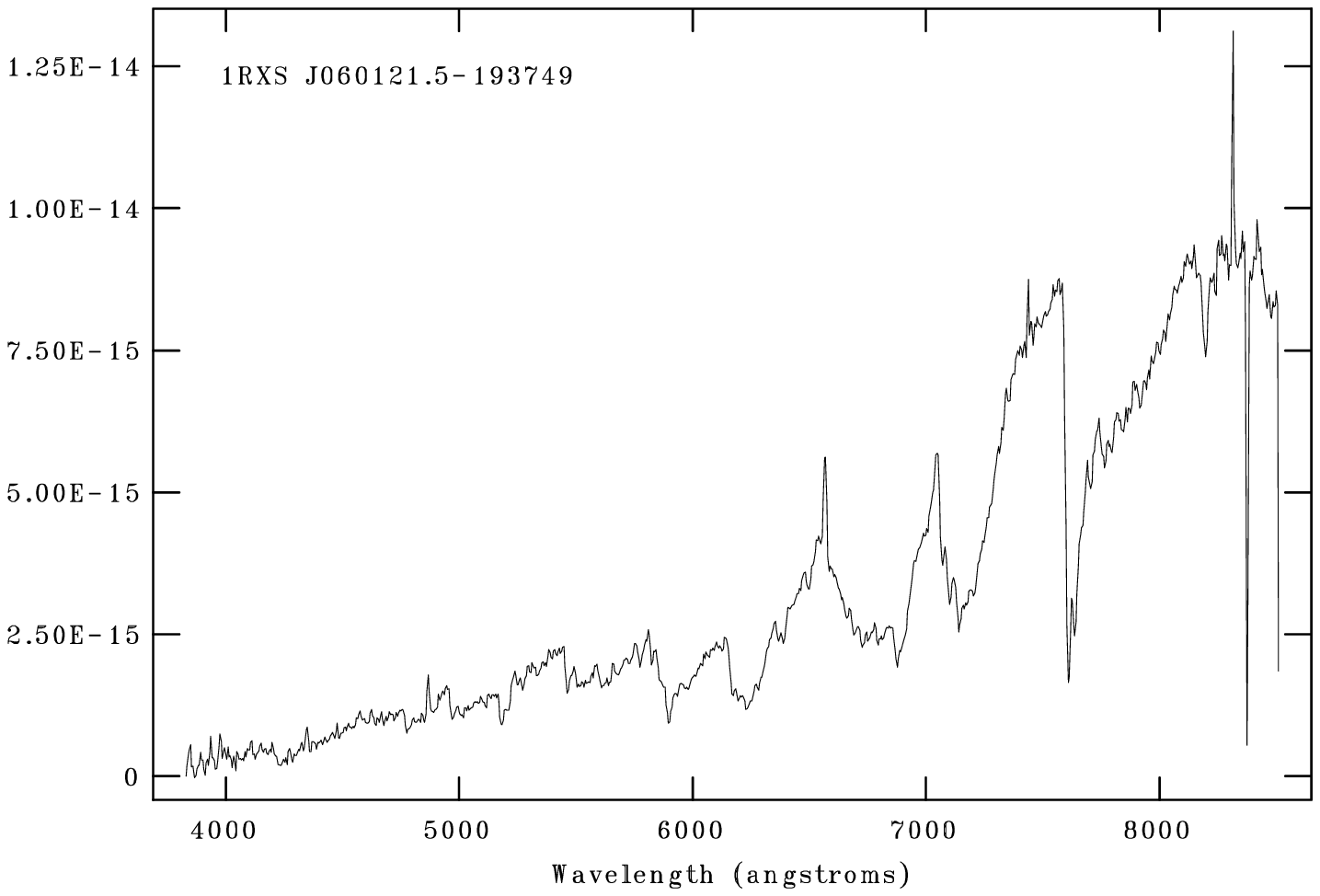}
\includegraphics[width=73mm]{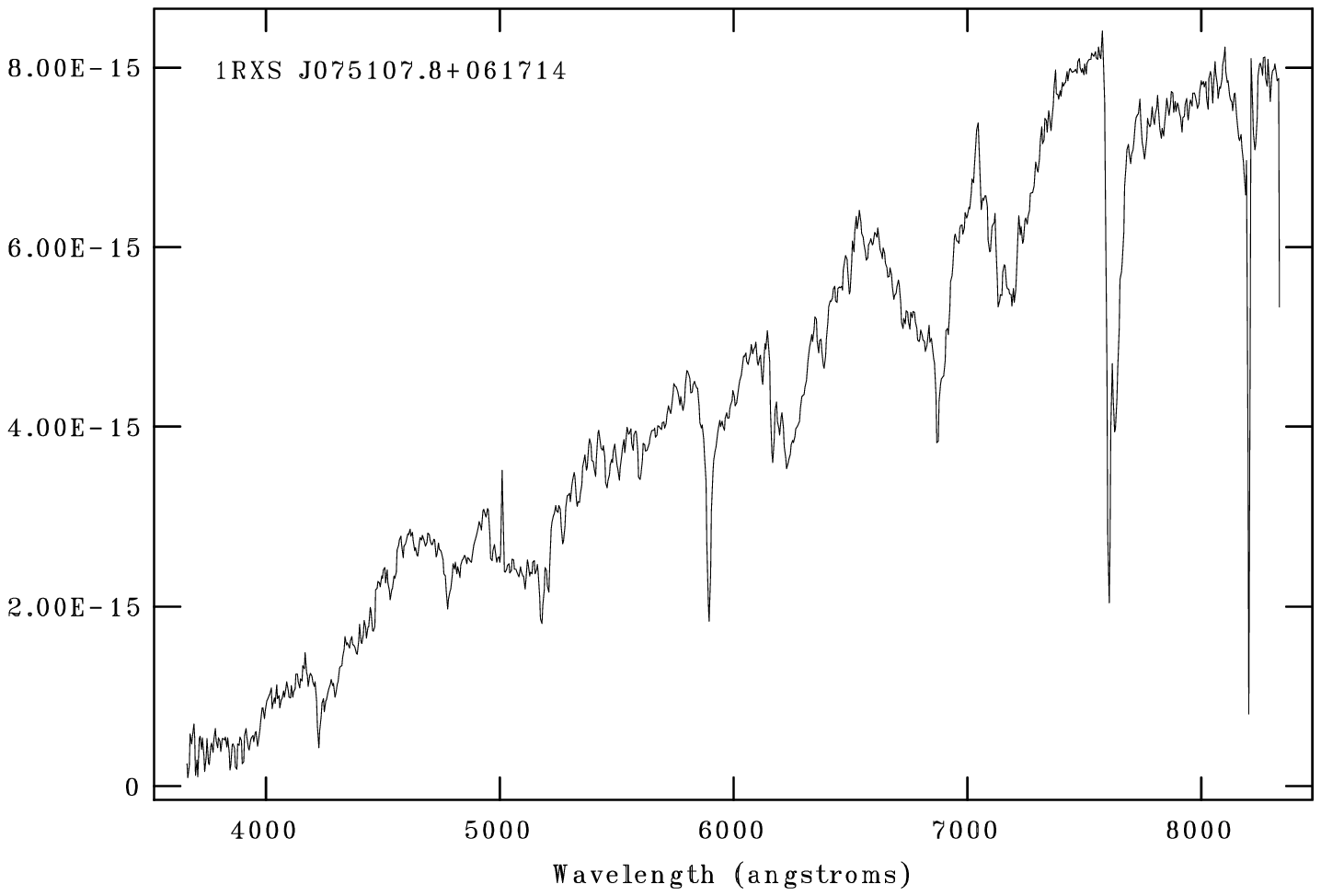}
\includegraphics[width=73mm]{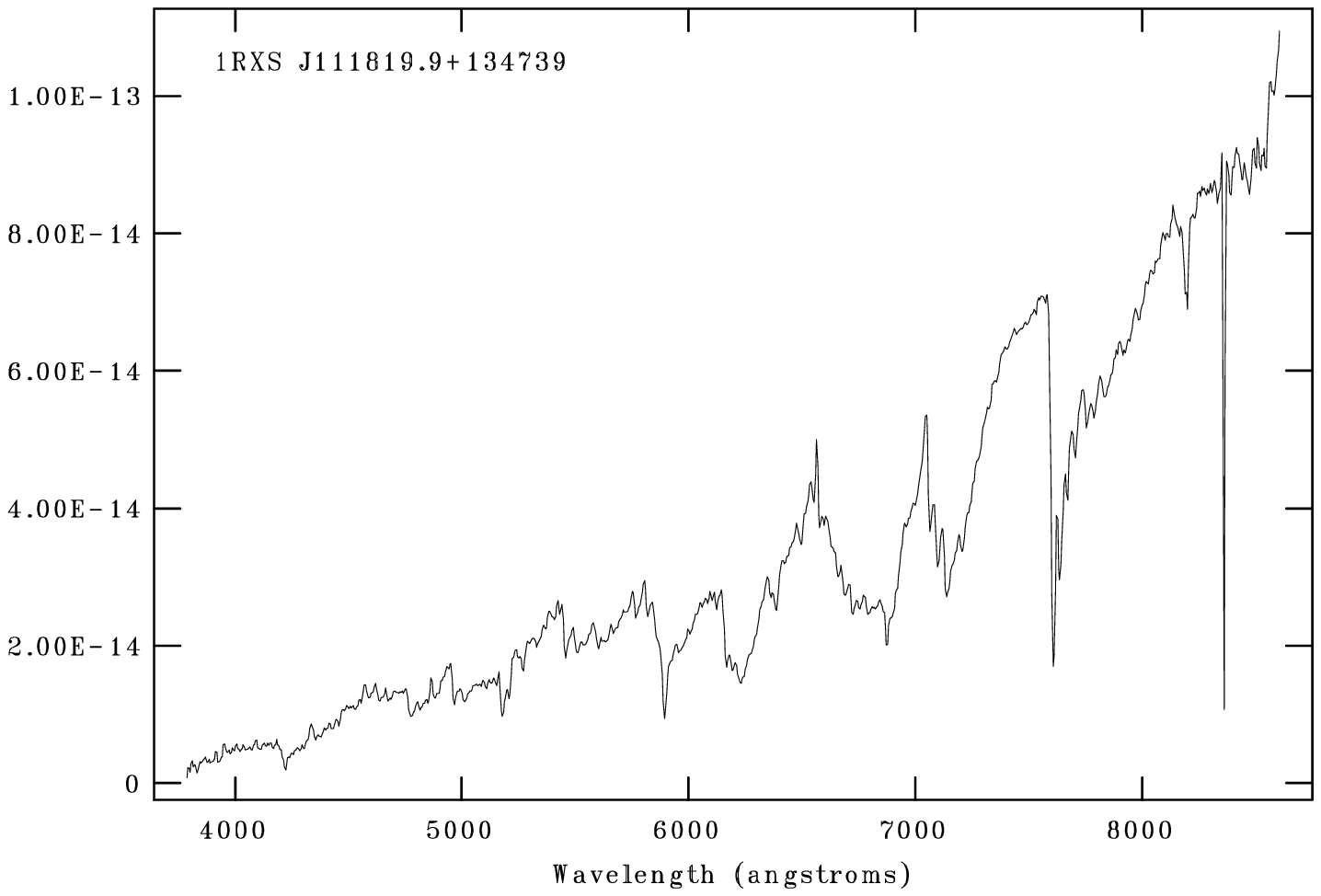}
\includegraphics[width=73mm]{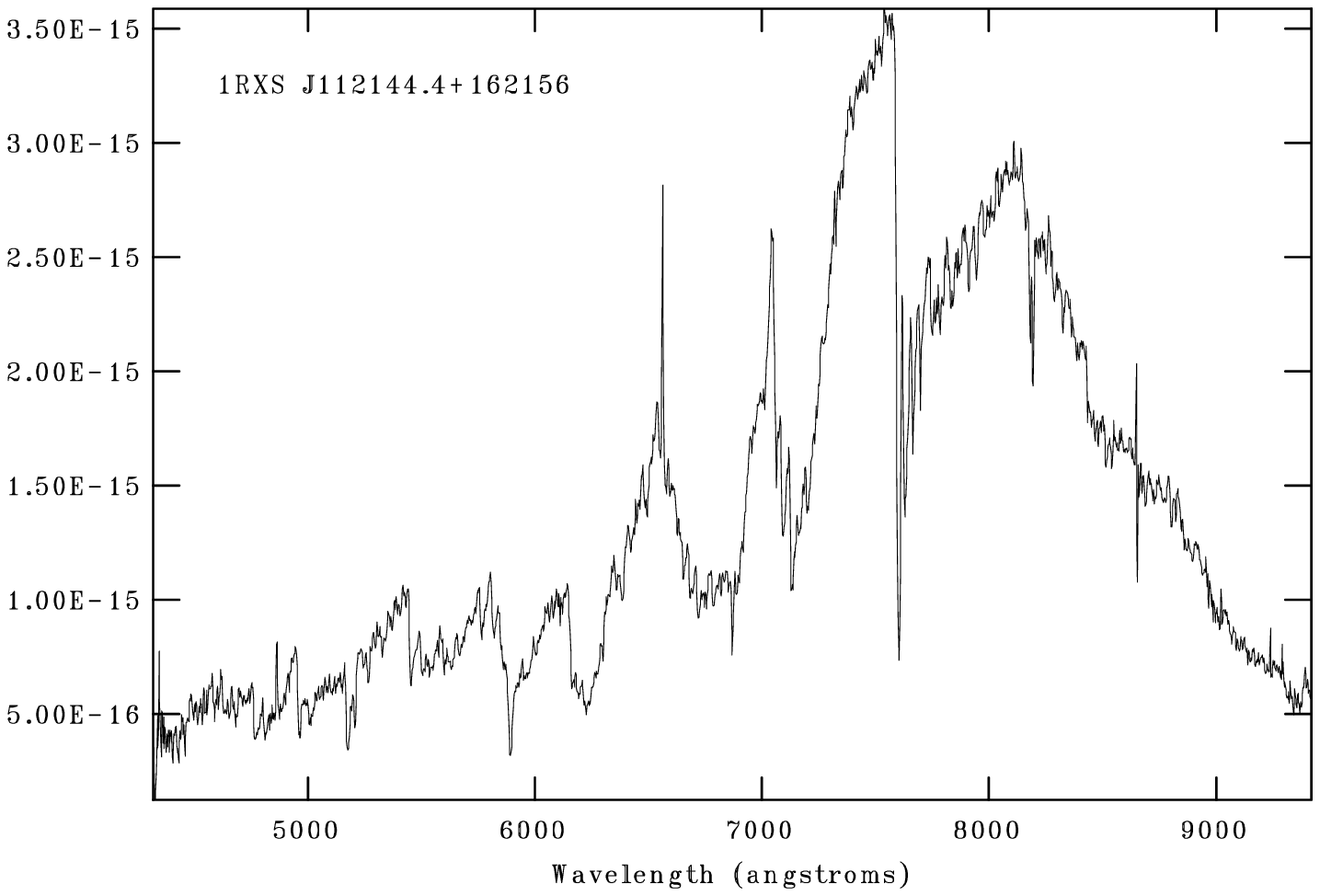}
\includegraphics[width=73mm]{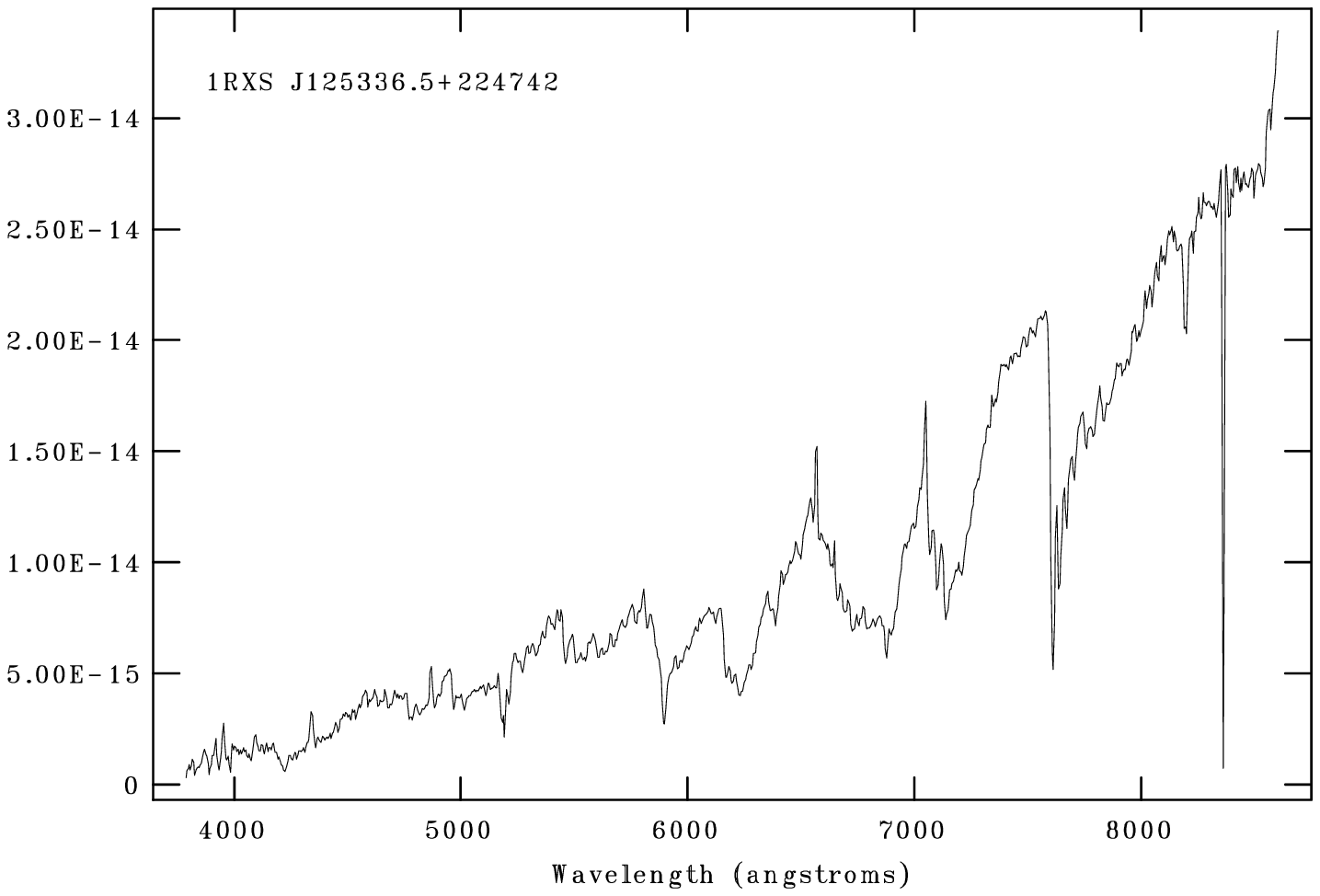}
\includegraphics[width=73mm]{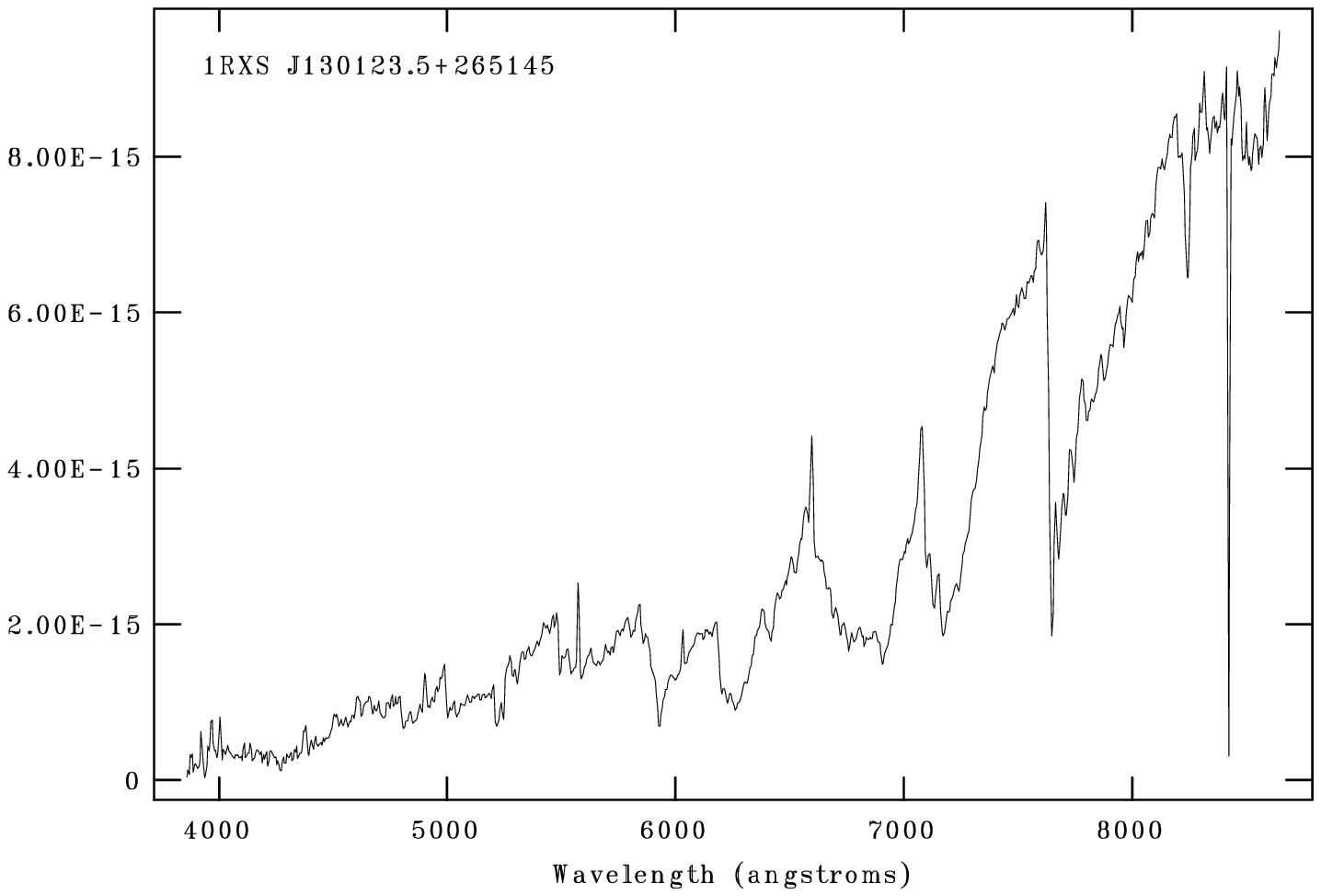}
\includegraphics[width=73mm]{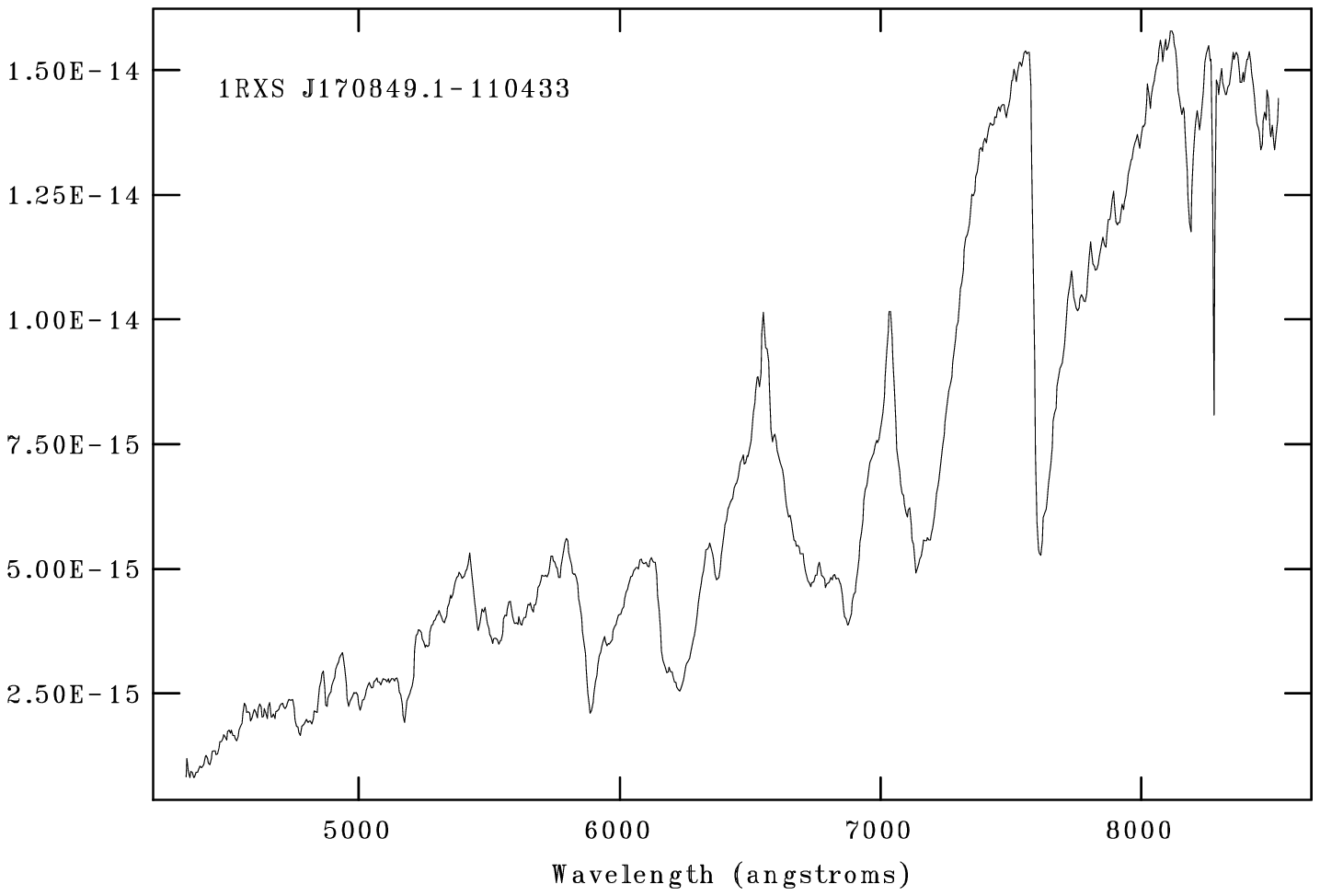}
\includegraphics[width=73mm]{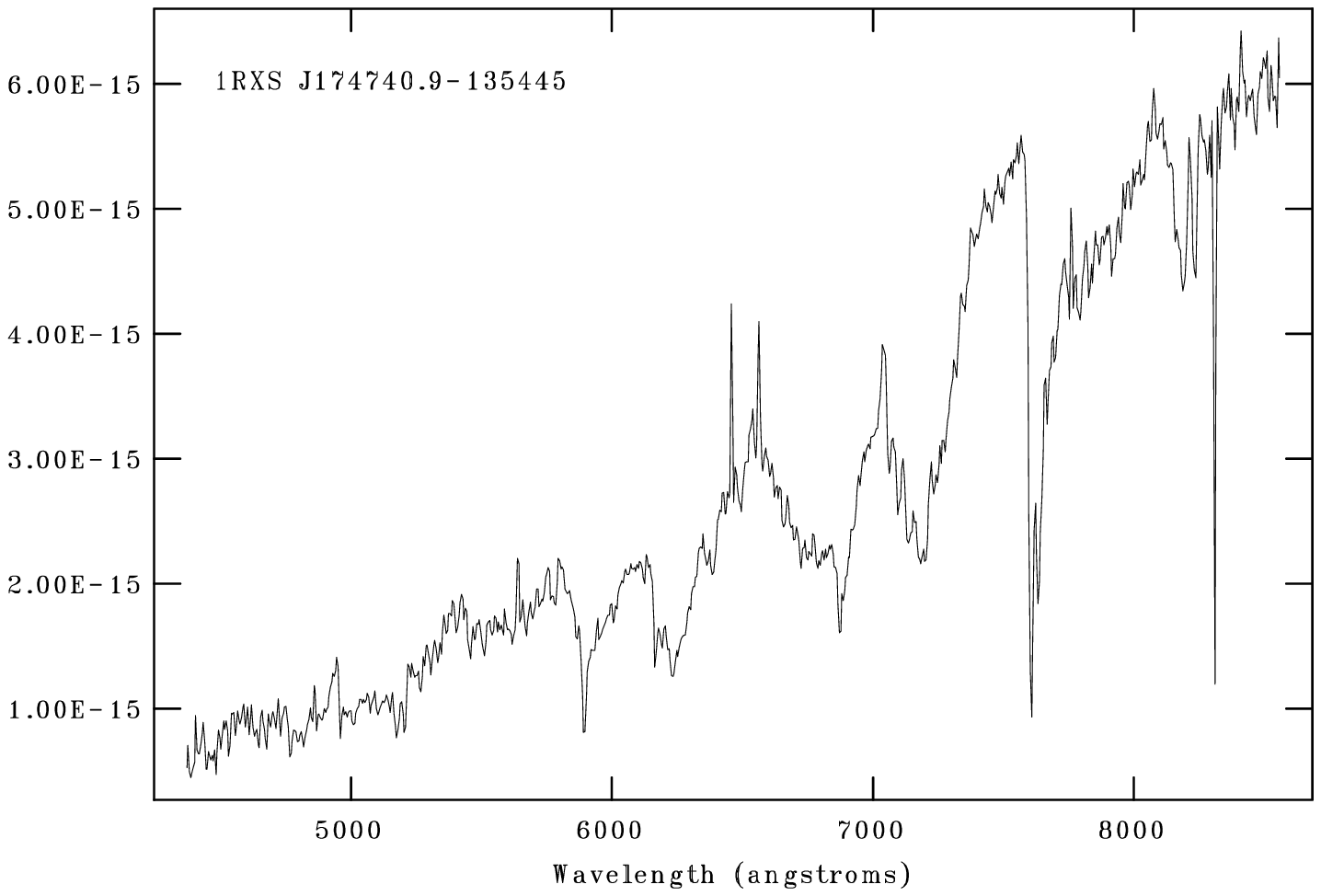}
\end{figure}
\begin{figure}
\includegraphics[width=73mm]{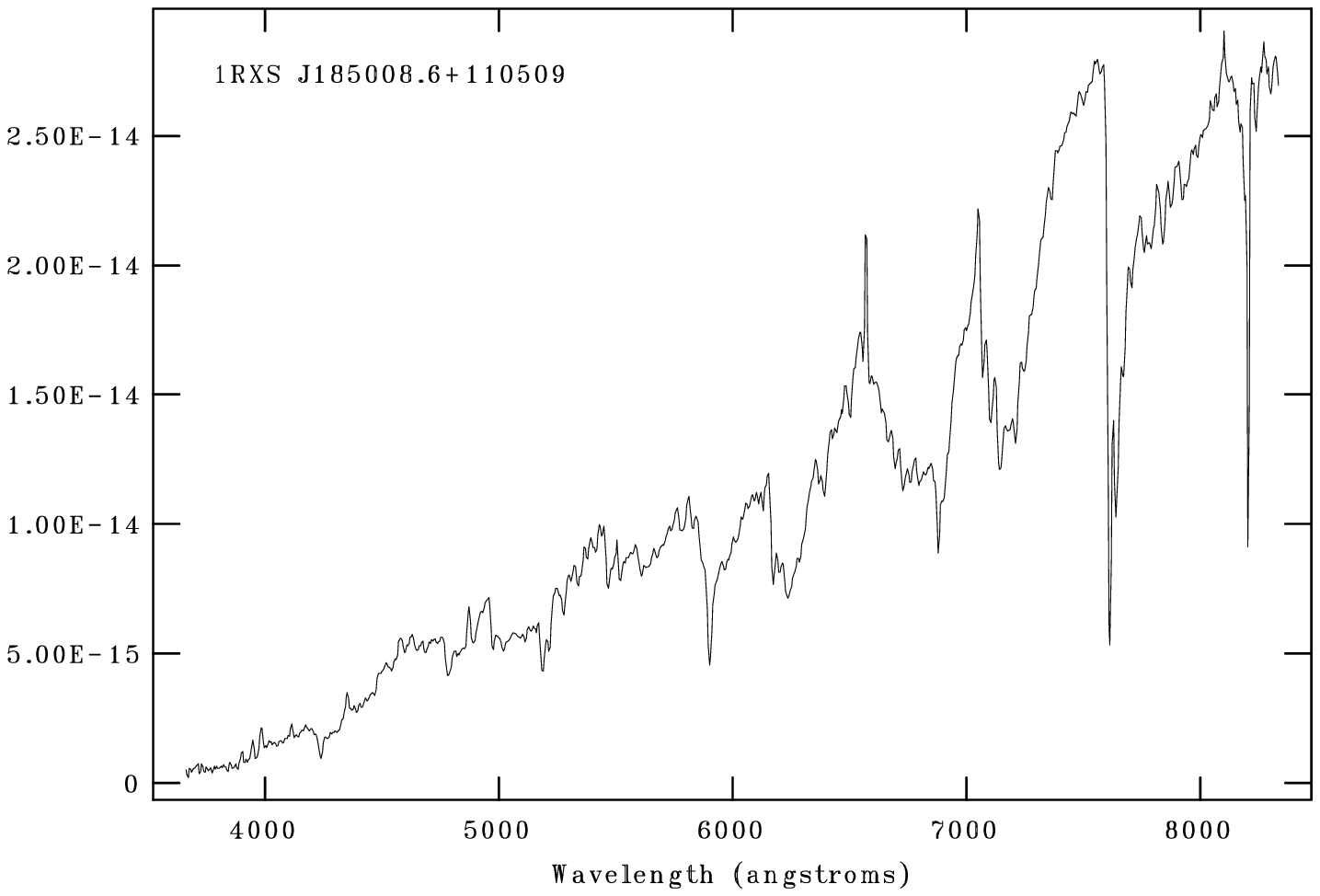}
\includegraphics[width=73mm]{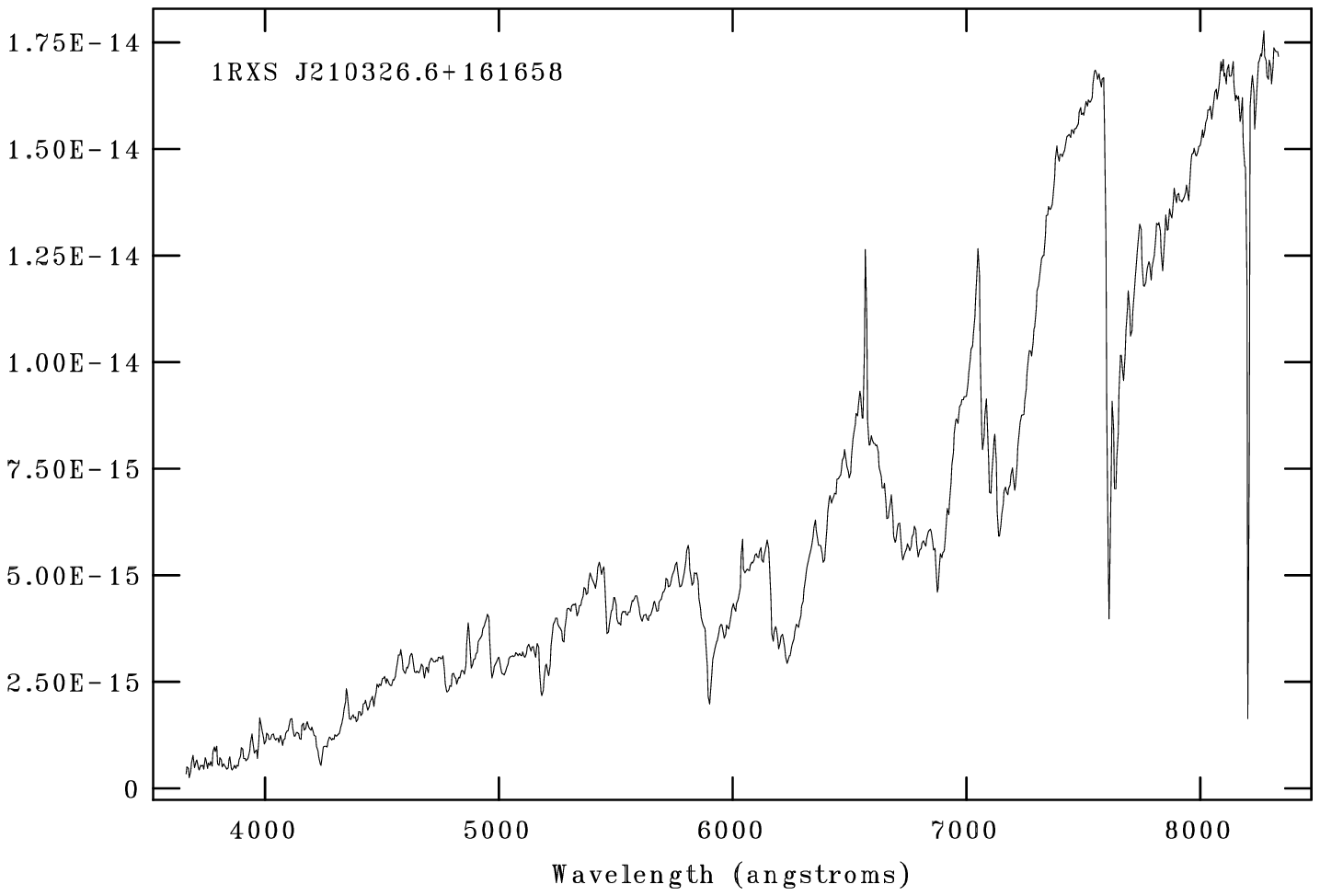}
\caption{Spectra of M dwarf stars. Spectral flux
($F_{\lambda}$) is in $\mathrm{erg\,cm^{-2}\,sec^{-1}\,\AA^{-1}}$. }
\label{spectra}
\end{figure}

\section{Discussion}
\subsection{Molecular Band Indices}
The classification of M dwarfs is based on the the
strength of the TiO and CaH molecular bands near
7000{\AA} and three spectral indices CaH2, CaH3,
and TiO5 are defined to measure the mean flux
level of the molecular bands (\cite{Reid95}).

In order to obtain the precise spectral indices, we
test wavelengths of $H_{\alpha}$ and find that
the wavelengths of the spectra shift up to
30{\AA} because of the long integral time of M dwarf
from 1200 to 3600 seconds. The gravity effect and the
change of temperature gradient may cause the
displacement between CCD and spectrograph in such a
long integral time. For each spectrum, the
central wavelength of $H_{\alpha}$ is measured
carefully to calculate the shift of molecular
band. The molecular band indices are displayed in
Table 2. There are discrepancies of molecular band
indices between our samples and six M dwarfs in
\cite{Riaz06}, which may be caused by the difference of
the spectral resolution. M dwarfs are by-product of
the MWQS and low resolution spectra (7-13\AA)
are good enough to make AGN identifications.

\begin{figure}
\includegraphics[width=73mm]{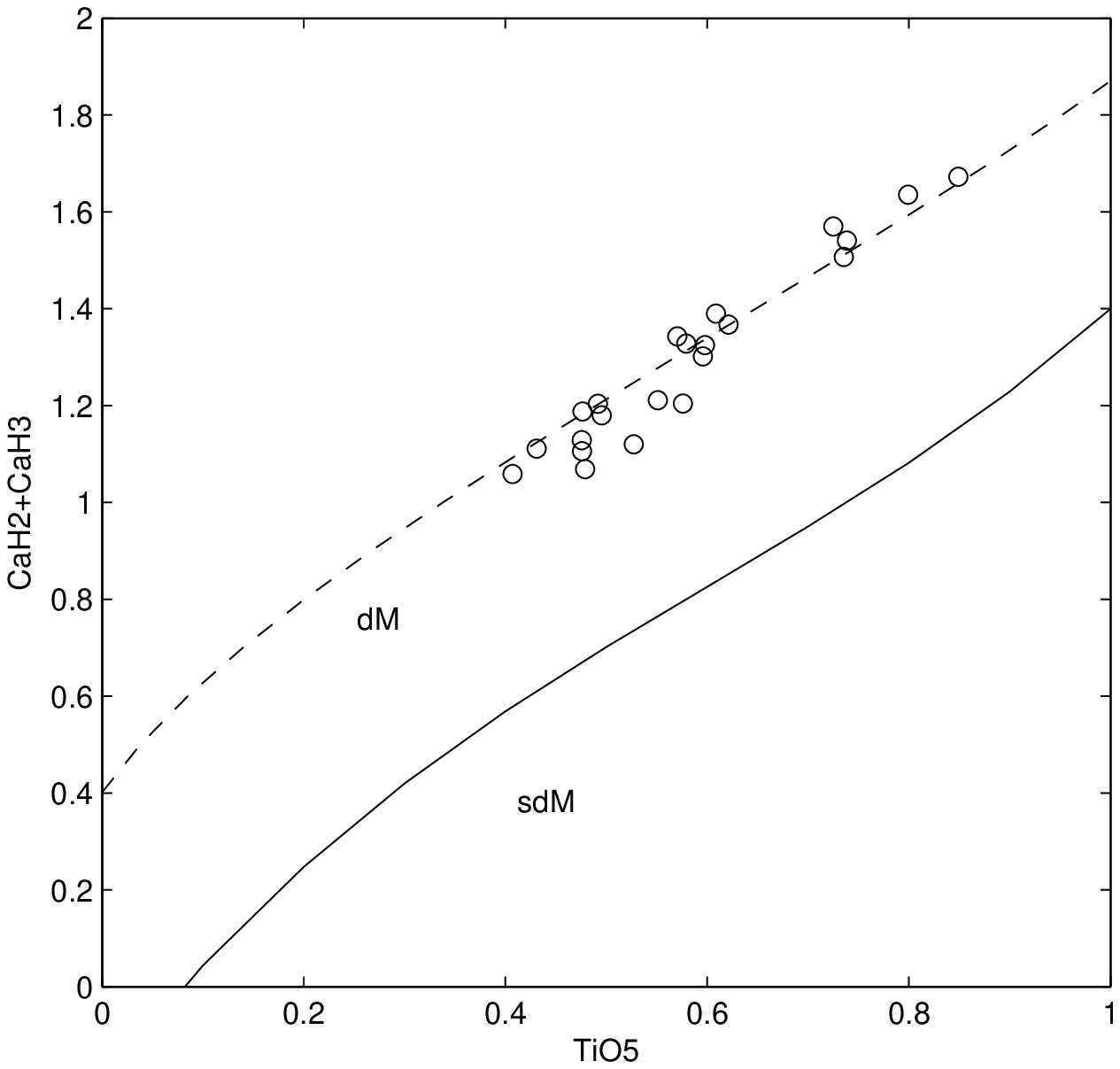}
\includegraphics[width=73mm]{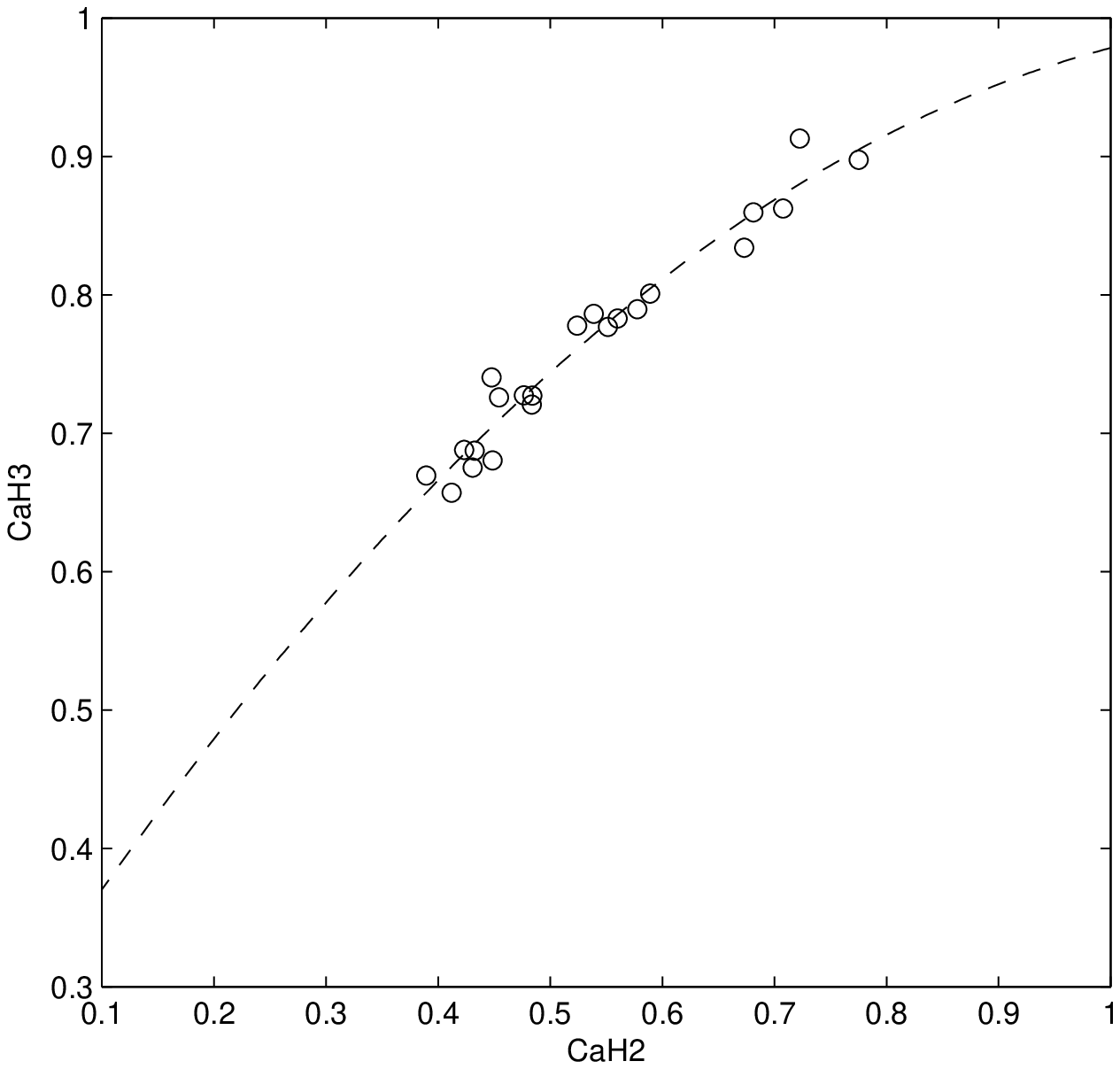}
\caption{{\it Left}: CaH and TiO band strengths of the stars of our sample.
{\it Dashed line} shows CaH2+CaH3/TiO5 relationship of disk M dwarfs
in \cite{Lepine07}. {\it Right}: Distribution of the CaH2 and CaH3 spectral
indices of our sample. {\it Dashed line} represents the best fitted relationship.}
\label{indices}
\end{figure}

The distribution of three spectral indices is
shown in Fig.~\ref{indices}. The TiO and CaH band
strengths are tightly correlated in the left frame,
which indicates that the stars have equivalent (solar)
metallicities and probably belong to Population I
(\cite{Lepine07}). The metallicity class
separators ({\it solid line}) proposed by
\cite{Burgasser06}, show that all our
stars belong to M dwarfs. The right frame in
Fig.~\ref{indices} represents the relation
between CaH2 and CaH3. The {\it Dashed line} stands
for the fitting of our sample with least square
method (Eq.~\ref{eq:CaH3}). The distribution
suggests that CaH2 may be dependent on
metallicity (\cite{Lepine07}).

\begin{equation}
CaH3=-0.516CaH2^{2}+1.244CaH2+0.251
\label{eq:CaH3}
\end{equation}

We obtain the parameter $\zeta_{TiO/CaH}$ introduced
by \cite{Lepine07} to make the metallicity class
separation among dwarfs, subdwarfs and extreme
subdwarfs, based on the relative strength of TiO
to CaH. The parameter $\zeta_{TiO/CaH}$ is defined as
\begin{equation}
\zeta_{TiO/CaH}=\frac{1-TiO5}{1-[TiO5]_{Z_{\odot}}},
\label{eq:zeta}
\end{equation}
where $[TiO5]_{Z_{\odot}}$ is given by \cite{Lepine07},
as a function of the CaH2+CaH3 index.
The parameters of our sample are lager
than 0.83 and confirm that they belong to M dwarfs
with near-solar metallicity.

We have used the relation from \cite{Reid95} (Eq.~\ref{eq:sp})
to obtain spectral type with an uncertainty of
$\pm$0.5 subclasses. The spectral type of our
sample ranges between K7 to M4.
\begin{equation}
{\it{S_{p}}}=-10.775\times TiO5+8.2
\label{eq:sp}
\end{equation}

\subsection{Absolute Magnitudes}

\cite{Riaz06} have given an experiential relation (Eq.~\ref{eq:MJ})
between $M_{J}$ and TiO5 for 35 M dwarfs with
TiO5$\geq$0.4, and an rms scatter is of 0.8 mag.
We used the relation to obtain $M_{J}$ ,
and derived the bolo-metric magnitudes from the
absolute K magnitudes (\cite{Veeder74}). The
distance of our stars was estimated by the
absolute magnitudes with the uncertainties on an
order of $\pm37\%$. There are 14 stars, over
half of the sample, which lie within 100 pc.
It has to be pointed out that stars that lie in
the region around TiO5$\sim$4 have a high uncertainty
in their absolute magnitudes and the derived distance
(\cite{Riaz06}).
\begin{equation}
M_{J}=0.97(TiO5)^{2}-5.01(TiO5)+8.73
\label{eq:MJ}
\end{equation}

\subsection{Luminosity}

We have used the relation of the count
rate to energy flux conversion (\cite{Schmitt95})
to obtain X-ray fluxes for our targets,
except one star with unreliable hardness ratio.
The X-ray luminosity was calculated with the
X-ray flux and the distance. The strong X-ray
luminosity of 20 targets with $L_x$ larger than
$10^{29}$, and one with
$L_x$ larger than $10^{28.5}$, suggests that they
have strong coronal activity (\cite{Barrado98}).

The distribution between Log($L_{x}/L_{bol}$) and
spectral type is shown in Fig.~\ref{X-ray
luminosity}. For a given spectral type, there is
a large spread in $L_{x}$, as shown in FIG. 5 of
\cite{Riaz06}. The rotational velocity
differences can explain this dispersion in $L_{x}$
for low-mass stars (\cite{Stauffer97}). The distribution
also suggests that samples with different distance
locate in the different areas of the diagram. Our
samples with distance exceeding 100pc incline to be
stronger X-ray emitters and earlier type stars than
ones within 100pc, but we need more samples to
confirm the trend. The distribution of
X-ray flux for dM and dMe stars is also tested
but no break is found.

\begin{figure}
\centering
\includegraphics[width=90mm]{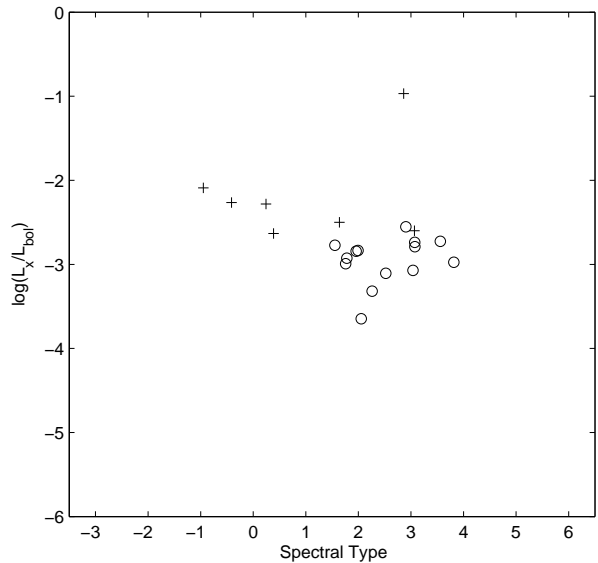}
\caption{Log($L_{x}/L_{bol}$) versus spectral type.
{\it Pluses} are targets with distance larger than 100pc and
{\it circles} stand for targets with distance less than 100pc.}
\label{X-ray luminosity}
\end{figure}

\subsection{Colors of Infrared}

\begin{figure}
\centering
\includegraphics[width=90mm]{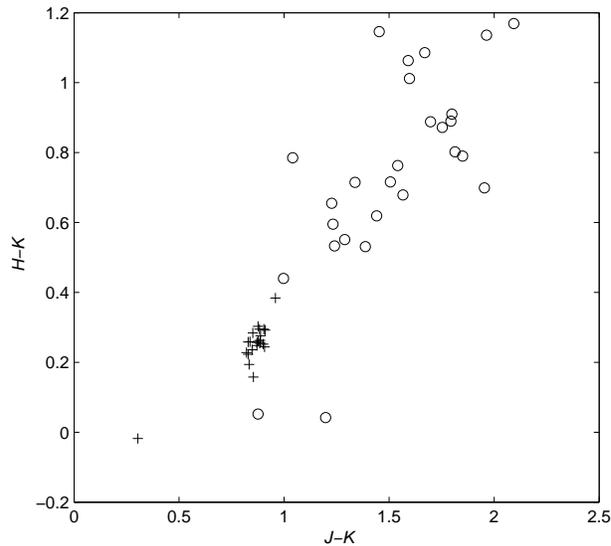}
\caption{Infrared colors diagram for M dwarfs and AGNs. {\it Pluses}: M dwarfs. {\it circles}: AGNs sample in \cite{Chen07}.}
\label{infrared colors}
\end{figure}

M dwarfs can not be distinguished from AGN
samples directly if criteria in soft X-ray band alone
are applied, but with colors in infrared band there
may be hints to differentiate the two kinds of sources.
Fig.~\ref{infrared colors} shows the color-color
diagram in infrared band for M dwarfs and AGN
samples (\cite{Chen07}). Colors of AGNs
distribute widely in the diagram because of the
complex hot dusty environment outside the
accretion disks with a distance of a few parsecs to
the central black holes. However, M dwarfs concentrate
in smaller area in the color-color diagram,
and this maybe indicate that the temperatures of
their atmosphere cover a narrow range from
$\sim$2100K to $\sim$3300K.

\cite{Chen07} suggested that most AGNs span a
color range $1.0<J-K<2.0$ and $0.5<H-K<1.2$,
while \cite{Riaz06} selected M dwarfs with
criteria $0.8<J-K<1.1$, $H-K>0.15$ and
$J-H<0.75$. Most of our targets are agreement
with the criteria of \cite{Riaz06}, but one M dwarf lies outside
the range. Color indices for our
sample are in ranges of about $0.76<J-K<0.96$, $H-K>0.14$, and 
$J-H<0.56$. The colors of infrared may be effective to 
distinguish M dwarfs from AGNs. 

\section{Summary}

As a by-product in MWQS, we have obtained the
spectra of 22 X-ray flux limited M dwarf stars.
The spectral types are calculated with molecular
band indices based on a calibration of the TiO to
CaH band strength ratio, ranged from K7 to M4.
Over half of them lie within 100pc of the Sun.
The distribution between Log($L_{x}/L_{bol}$) and
spectral type suggests that M dwarfs with
different distant may locate in different areas
of the diagram and the large spread in X-ray
luminosities for a given spectral type reflects
the range in the rotational velocities of the
stars. The colors of infrared could be criteria
to separate AGNs from M dwarfs selected from soft
X-ray data. For the future observation, we are
going to enlarge the samples up to 100 and a
further investigation on X-ray luminosity
function may provide more information about
coronal heating mechanisms and occurrence of
flares. An observational campaign is on the
schedule aimed at obtaining frequency spectrum
of dM and dMe stars based on photometric
observation to characterize M dwarfs variations,
which will give us a chance to study the interior
pulsation models of M dwarfs. Extrasolar planet
candidates may be selected if we can get long term
photometric observation and high resolution spectral
observation.

\normalem
\begin{acknowledgements}
This work was supported by Scientific Research Foundation of Beijing Normal
University and the National Natural Science Foundation of China (NSFC)
under No.10778717. This work was partially Supported by the Open Project
Program of the Key Laboratory of Optical Astronomy, National Astronomical
Observatories, Chinese Academy of Sciences.
\end{acknowledgements}

\begin{table}[]
\centering
\rotcaption{Quantities of M Dwarf Stars}
\begin{sideways}
\begin{threeparttable}



%
 \begin{tabular}{lrrrrrrrrclrr}
  \hline\noalign{\smallskip}
ROSAT name            & cts\tnote{a}  &  HR1\tnote{b}  & J      & H      & K      &  CaH2  &   CaH3  &   TiO5  &$\zeta_{TiO/CaH}$& SpT\tnote{c}    &     Distance\tnote{d}  & log($L_{x}/L_{bol}$) \\
  \hline\noalign{\smallskip}
1RXS J015515.4-173916 & 0.0622 & -0.23 & 12.268 & 11.645 & 11.397 &  0.72  &   0.91  &   0.80  &   1.19 & M0     &      242 &   -2.26      \\
1RXS J040840.7-270534 & 0.0610 & -0.08 & 10.813 & 10.234 &  9.975 &  0.42  &   0.69  &   0.43  &   0.98 & M3.5   &       64 &   -2.73      \\
1RXS J041132.8+023559 & 0.0825 &  0.06 & 10.416 &  9.793 &  9.530 &  0.58  &   0.78  &   0.62  &   1.01 & M1.5   &       76 &   -2.77      \\
1RXS J041132.8+023559 & 0.0825 &  0.06 & 10.046 &  9.428 &  9.136 &  0.52  &   0.78  &   0.60  &   0.94 & M2     &       61 &   -2.93      \\
1RXS J041325.8-013919 & 0.0976 &  0.01 &  9.375 &  8.761 &  8.504 &  0.43  &   0.69  &   0.53  &   0.83 & M2.5   &       40 &   -3.11      \\
1RXS J041612.7-012006 & 0.0358 & -0.26 & 12.004 & 11.429 & 11.045 &  0.45  &   0.74  &   0.48  &   1.01 & M3     &      122 &   -2.60      \\
1RXS J041615.7+012640 & 0.0211 &       & 13.602 & 12.962 & 12.768 &  0.67  &   0.83  &   0.74  &   0.99 & M0.5   &      403 &              \\
1RXS J042854.3+024836 & 0.0203 & -1.00 & 10.809 & 10.203 &  9.980 &  0.56  &   0.78  &   0.57  &   1.08 & M2     &       83 &   -3.65      \\
1RXS J043051.6-011253 & 0.0172 &  0.21 & 11.403 & 10.830 & 10.527 &  0.39  &   0.67  &   0.41  &   0.96 & M4     &       81 &   -2.98      \\
1RXS J043426.2-030041 & 0.0325 &  0.15 & 13.141 & 12.445 & 12.287 &  0.78  &   0.90  &   0.85  &   1.08 & K7     &      391 &   -2.09      \\
1RXS J053954.8-130805 & 0.0505 & -0.24 & 10.601 &  9.984 &  9.724 &  0.54  &   0.79  &   0.60  &   0.98 & M2     &       80 &   -2.99      \\
1RXS J055533.1-082915 & 0.0717 & -0.20 & 10.735 & 10.168 &  9.884 &  0.43  &   0.68  &   0.48  &   0.90 & M3     &       68 &   -2.74      \\
1RXS J060121.5-193749 & 0.0956 & -0.01 & 14.104 & 13.782 & 13.800 &  0.45  &   0.73  &   0.50  &   0.96 & M3     &      333 &   -0.97      \\
1RXS J075107.8+061714 & 0.0696 &  0.11 & 11.866 & 11.257 & 10.962 &  0.68  &   0.86  &   0.74  &   1.08 & M0     &      182 &   -2.28      \\
1RXS J111819.9+134739 & 0.1086 & -0.47 &  9.087 &  8.516 &  8.258 &  0.48  &   0.73  &   0.55  &   0.90 & M2.5   &       36 &   -3.32      \\
1RXS J112144.4+162156 & 0.0278 & -0.39 & 12.079 & 11.450 & 11.195 &  0.71  &   0.86  &   0.73  &   1.26 & M0.5   &      196 &   -2.63      \\
1RXS J125336.5+224742 & 0.0758 & -0.12 & 10.482 &  9.870 &  9.634 &  0.45  &   0.68  &   0.48  &   0.93 & M3     &       60 &   -2.79      \\
1RXS J130123.5+265145 & 0.0136 &  0.30 & 11.333 & 10.740 & 10.512 &  0.41  &   0.66  &   0.48  &   0.85 & M3     &       90 &   -3.07      \\
1RXS J170849.1-110433 & 0.0769 & -0.25 & 10.541 &  9.953 &  9.658 &  0.48  &   0.72  &   0.58  &   0.84 & M2     &       74 &   -2.84      \\
1RXS J174741.0-135445 & 0.0552 &  1.00 & 11.058 & 10.395 & 10.151 &  0.59  &   0.80  &   0.61  &   1.09 & M1.5   &      100 &   -2.50      \\
1RXS J185008.6+110509 & 0.0556 &  0.43 & 10.429 &  9.817 &  9.541 &  0.55  &   0.78  &   0.58  &   1.03 & M2     &       71 &   -2.84      \\
1RXS J210326.6+161658 & 0.0750 & -0.02 & 11.078 & 10.430 & 10.178 &  0.48  &   0.73  &   0.49  &   1.00 & M3     &       82 &   -2.55      \\
  \noalign{\smallskip}\hline

\end{tabular}
\begin{tablenotes}
\item[a]The {\it{ROSAT}} count rate in counts s$^{-1}$.
\item[b]The {\it{ROSAT}} hardness ratio, except one star with unreliable hardness ratio.
\item[c]Spectral type.
\item[d]Distance in parsec.
\end{tablenotes}
\end{threeparttable}
\end{sideways}

\end{table}


\end{document}